\documentclass[11pt,letterpaper]{article}
\usepackage[margin=1in]{geometry}
\usepackage[T1]{fontenc}
\usepackage{lmodern}
\usepackage{amsmath,amssymb,amsthm,mathtools}
\usepackage{microtype}
\usepackage{booktabs,array,graphicx}
\usepackage{tikz}
\usetikzlibrary{arrows.meta,calc,decorations.pathreplacing,positioning}
\usepackage{float}
\floatstyle{ruled}
\newfloat{algorithm}{tbp}{loa}
\floatname{algorithm}{Algorithm}
\usepackage{xurl}
\usepackage[unicode,hidelinks]{hyperref}
\hypersetup{pdftitle={Rounded Potentials Certify Exact Maximum Flow},pdfauthor={Ohad Trabelsi}}
\usepackage[nameinlink,noabbrev]{cleveref}
\newcounter{certificatebox}
\crefname{certificatebox}{Box}{Boxes}
\Crefname{certificatebox}{Box}{Boxes}
\usepackage{enumitem}
\setlist[itemize]{topsep=3pt,itemsep=1pt}
\setlist[enumerate]{topsep=3pt,itemsep=2pt}
\allowdisplaybreaks
\newcommand{\tO}{\widetilde O}
\newcommand{\ex}{\operatorname{ex}}
\newcommand{\clip}{\operatorname{clip}}
\newcommand{\MaxFlow}{\operatorname{MaxFlow}}
\newcommand{\one}{\mathbf{1}}
\newcommand{\evec}{\mathbf{e}}
\newcommand{\MM}{\operatorname{MM}}
\newtheorem{theorem}{Theorem}[section]
\newtheorem{observation}[theorem]{Observation}
\newtheorem{corollary}[theorem]{Corollary}
\newtheorem{lemma}[theorem]{Lemma}
\theoremstyle{definition}
\newtheorem{definition}[theorem]{Definition}
\newtheorem{example}[theorem]{Example}
\theoremstyle{remark}
\newtheorem{remark}[theorem]{Remark}
\theoremstyle{plain}
\title{Rounded Potentials Certify Exact Maximum Flow}
\author{Ohad Trabelsi\\University of Haifa, Israel}
\date{November 15, 2026}
\begin{document}
\maketitle
\begin{abstract}
\small
\setlength{\emergencystretch}{1em}
The textbook certificate for maximum flow is lopsided: the cut takes one bit per vertex, while the flow takes one number per edge. Indeed, some graphs have $\Theta(n^2)$ terminal pairs for which every maximum flow uses $\Theta(n^2)$ edges. Yet we show that one short number per vertex suffices for both cut and flow:
$n$ vertex potentials, each rounded to $O(\log(nW))$ bits, certify the exact maximum $st$-flow value in any directed graph with integer capacities at most $W$.
Exact potentials can have exponentially large representations. Rounding them may violate flow conservation, but the total error across any cut is less than one unit, so integrality of cut capacities absorbs it.

Checking conservation is local and linear in the capacities, allowing the same certificate to yield new bounds in four models (capacities are polynomially bounded unless stated otherwise):
\begin{itemize}
\item \textbf{Online annotated streaming.} $\widetilde{O}(n)$ total cost (help plus space) for the maximum $st$-flow value, optimal up to logarithmic factors, answering a question of Thaler (ICALP 2016) on whether $o(n^2)$ is possible.
\item \textbf{Broadcast proof-labeling.} $O(\log(n+p))$-bit labels certifying a maximum $st$-flow value of $p$ in directed and undirected graphs with \textit{arbitrary} capacities, optimal for $p\ge n^{\Omega(1)}$, settling that regime of a question of Korman, Kutten, and Peleg (Distributed Computing 2010).
\item \textbf{Merlin--Arthur verification.} $\widetilde{O}(n^3)$ time and $O(n^3\log n)$ proof length for directed All-Pairs Max-Flow, improving on previous verification times of $n^{4+o(1)}$ for general capacities and $n^{3.5+o(1)}$ for unit capacities, and answering a question of Akmal (TheoretiCS 2024).
\item \textbf{Nondeterministic algorithms.} $O(n^{3.614})$ time for directed All-Pairs Max-Flow, or $n^{3.5+o(1)}$ if $\omega=2$: the first subquartic bound for general edge capacities, answering a question of Trabelsi (SODA 2025).
\end{itemize}
\end{abstract}
\clearpage

\tableofcontents
\clearpage

\section{Introduction}
The max-flow/min-cut theorem of Ford and Fulkerson~\cite{FF56} can also be viewed as a statement about certificates. A feasible $st$-flow of value $p_1$ shows that the maximum flow value is at least $p_1$; an $st$-cut of capacity $p_2$ shows that the minimum cut capacity is at most $p_2$; and the theorem guarantees that both witnesses can be chosen with $p_1=p_2$. Together they form the classical certificate for the claim $\MaxFlow(s,t)=p$, and one can verify it in linear time. The two parts, however, may differ significantly in size. The cut requires $n$ bits, one per vertex, while the flow requires one number per edge. There are graphs in which, for $\Theta(n^2)$ pairs $(u,v)$, every maximum $uv$-flow uses $\Theta(n^2)$ edges (see Observation~\ref{obs:dense}). Explicit flow witnesses therefore have $\Theta(n^2)$ entries for one pair and $\Omega(n^4)$ entries altogether. In such instances, the flow dominates the certificate size.

Several settings admit flow witnesses that avoid explicit description. In undirected unit-capacity graphs, tree packings~\cite{BFJ95,CH03} witness all single-source lower bounds at once~\cite{AKT21}. This ingredient underlies the near-linear-time verification of Gomory--Hu trees in~\cite{AKT21}. A Gomory--Hu tree~\cite{GH61} encodes a minimum cut for every pair of an undirected graph using a tree on $n-1$ edges, but by itself certifies only the cut upper bounds; the matching flows need separate justification. For planar digraphs, \L{}\k{a}cki, Nussbaum, Sankowski, and Wulff-Nilsen~\cite{LNSW12} give implicit maximum-flow representations based on dual distances and face potentials.
For directed graphs with polynomially bounded integer capacities, van den Brand, Song, and Weng~\cite{vdBSW26} compute the exact maximum-flow value using a randomized streaming algorithm with multiple passes and $\widetilde O(n^{3/2})$ space. Their algorithm stores a finite-precision sequence of augmenting circulations using vertex potentials and interior-point auxiliary sketches. An edge amount can be approximately recovered from this representation together with the queried edge and its attributes.

We seek succinct, deterministically verifiable maximum-flow certificates. Equivalently, a nondeterministic algorithm guesses a certificate and verifies the claimed value. We ask:
\begin{center}
\emph{How succinctly can we certify maximum flow in general graphs?}
\end{center}

A natural approach is to represent a maximum flow using vertex potentials. Consider a maximum flow minimizing the quadratic energy $\sum_{e:c_e>0} f_e^2/(2c_e)$. Classical optimality conditions for such convex-cost flows~\cite{Roc84,AMO93} give a suitable representation. However, the exact representation need not be succinct: even with unit capacities, some edge amounts in the minimum-energy flow can have denominators exponentially large in $n$ (see Subsection~\ref{sec:illustrations}). We show that $n$ vertex potentials, each rounded to $O(\log(nW))$ bits, suffice to certify the exact maximum-flow value. The potentials define a capacity-feasible pseudoflow and encode an $st$-cut, but rounding may destroy flow conservation.

\begin{figure}[p]
\noindent\begingroup\setlength{\fboxsep}{8pt}\fbox{\begin{minipage}{\dimexpr\linewidth-2\fboxsep-2\fboxrule\relax}
\refstepcounter{certificatebox}\label{box:certificate}%
\textbf{Box~\thecertificatebox. An exact maximum-flow certificate.}
\par\smallskip\noindent
Let capacities $c_{uv}\le W$ and the claimed value $p\ge0$ be integers. The \emph{target excess} is $b(s)=p$, $b(t)=-p$, and $b(v)=0$ for every $v\in V\setminus\{s,t\}$.

\par\medskip\noindent
\textbf{The certificate.} Let $Q_0=16n^2W$, and guess potentials $y_v\in[0,n-1]\cup[n,2n-1]$ that are multiples of $1/Q_0$, with $y_s\ge n$ and $y_t\le n-1$. Each potential uses $O(\log(nW))$ bits. The implied pseudoflow $f$ need not conserve flow exactly; the verification allows error $1/(8n)$ at each vertex. Define
\begin{equation}\label{eq:clip}
f_{uv}=c_{uv}\clip(y_u-y_v,0,1),\qquad
\clip(z,a,b)=\min\{b,\max\{a,z\}\}\quad(a\le b).
\end{equation}
Let $\ex_f(v)=\sum_u(f_{vu}-f_{uv})$ for the net outflow at $v$. Check that
\[
|\ex_f(v)-b(v)|\le1/(8n)
\quad\text{at every vertex.}
\]
\par\medskip\noindent
\textbf{Correctness (the value is exact).}\par\smallskip\noindent
\emph{Every cut has capacity at least $p$.} In any $st$-cut $(A,V\setminus A)$, the target excesses sum to $p$. The triangle inequality gives
\[
\left|\sum_{v\in A}\ex_f(v)-p\right|\le\frac{|A|}{8n}<1.
\]
Internal edges cancel in the sum, leaving the net flow out of $A$. The capacity of the cut is at least this net outflow, so integrality makes every $st$-cut's capacity at least $p$.
\par\smallskip\noindent
\emph{The encoded cut has capacity exactly $p$.} Set $S=\{v:y_v\ge n\}$, an $st$-cut. Every potential in $S$ exceeds every potential in $V\setminus S$ by at least one, so all outgoing cut edges are full and all incoming edges are empty. The cut capacity equals net outflow, so the same bound forces its integer capacity to be exactly $p$. Max-flow/min-cut gives $\MaxFlow(s,t)=p$.
\par\medskip\noindent
\textbf{Existence (such potentials exist).} Minimizing $\sum_{e:c_e>0}f_e^2/(2c_e)$ over maximum flows gives initial potentials $x\in[0,n-1]^V$ satisfying~\eqref{eq:clip} (Lemma~\ref{lem:potentials}). Choose a minimum cut $S$ and set
\[
y_v=x_v+n\one[v\in S].
\]
Internal drops are unchanged; outgoing cut edges were already full and incoming ones empty, so the shift preserves every edge amount and puts the potentials in the required intervals. Round these potentials down to multiples of $1/Q_0$. The rounding errors lie in $[0,1/Q_0)$, so potential differences change by less than $1/Q_0$. Clipping is $1$-Lipschitz, giving edge errors $\le c_{uv}/Q_0$ and excess errors $\le2(n-1)W/Q_0<1/(8n)$. Interval endpoints are grid points, so rounding preserves each potential's interval membership.

\smallskip
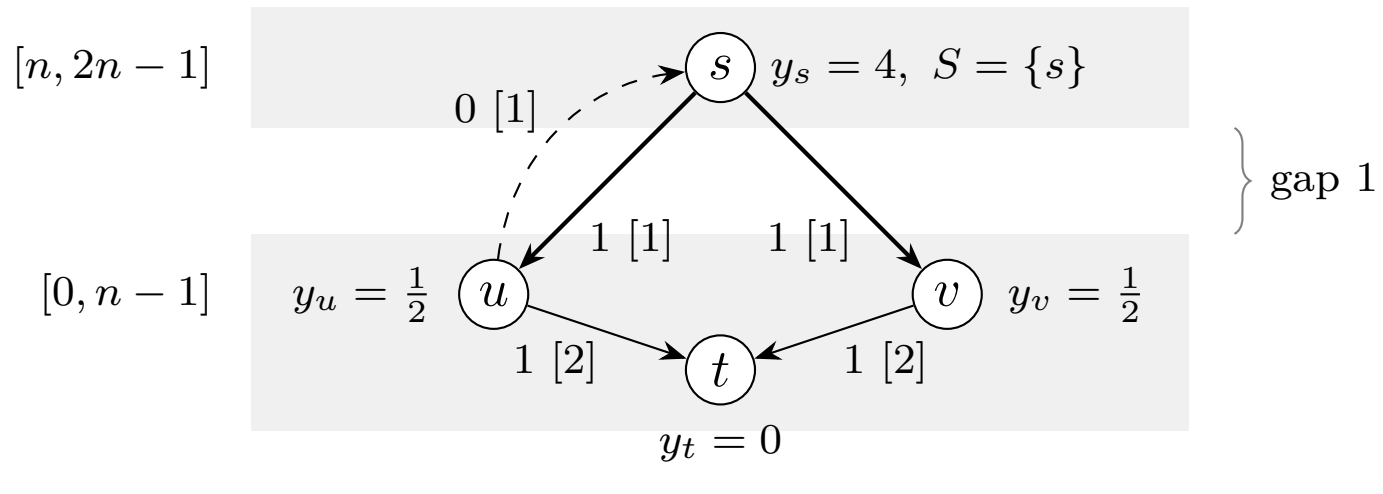
\begin{center}
\begin{tikzpicture}[x=1cm,y=1cm,
    vtx/.style={circle,draw,fill=white,inner sep=0pt,minimum size=13pt,font=\small},
    arc/.style={-{Stealth[length=5pt]}},
    lab/.style={font=\scriptsize,inner sep=1.5pt}]
  % the two permitted potential intervals and their unit separation
  \fill[gray!12] (.4,1.75) rectangle (6.6,2.55);
  \fill[gray!12] (.4,-.25) rectangle (6.6,1.05);
  \node[lab,anchor=east] at (.2,2.15) {$[n,2n-1]$};
  \node[lab,anchor=east] at (.2,.65) {$[0,n-1]$};
  \draw[gray,decorate,decoration={brace,amplitude=3pt}]
    (6.9,1.75) -- node[lab,right=5pt,text=black] {gap $1$} (6.9,1.05);
  % vertices grouped by their potential interval
  \node[vtx] (s) at (3.5,2.15) {$s$};
  \node[vtx] (u) at (2,.65) {$u$};
  \node[vtx] (v) at (5,.65) {$v$};
  \node[vtx] (t) at (3.5,.15) {$t$};
  \node[lab,anchor=west] at ([xshift=1pt]s.east) {$y_s=4,\ S=\{s\}$};
  \node[lab,anchor=east] at ([xshift=-3pt]u.west) {$y_u=\tfrac12$};
  \node[lab,anchor=west] at ([xshift=3pt]v.east) {$y_v=\tfrac12$};
  \node[lab,anchor=north] at ([yshift=-2pt]t.south) {$y_t=0$};
  % edges, labeled amount [capacity]
  \draw[arc,thick] (s) -- node[lab,pos=0.66,anchor=north west] {1 [1]} (u);
  \draw[arc,thick] (s) -- node[lab,pos=0.66,anchor=north east] {1 [1]} (v);
  \draw[arc] (u) -- node[lab,pos=0.5,anchor=north east] {1 [2]} (t);
  \draw[arc] (v) -- node[lab,pos=0.5,anchor=north west] {1 [2]} (t);
  \draw[arc,dashed] (u) to[bend left=38] node[lab,pos=0.45,anchor=south east] {0 [1]} (s);
\end{tikzpicture}

\par\smallskip
{\normalsize Flow [capacity], $n=4$, cut capacity $2$.\par}
\end{center}
\end{minipage}}
\par\endgroup
\end{figure}

In~\cite{vdBSW26}, integrality recovers the exact maximum-flow value from an approximation whose accuracy is guaranteed by the algorithm. Here, an untrusted prover supplies the potentials, so no such guarantee can be assumed. Instead, the verifier checks local conservation errors in the pseudoflow, establishing that every $st$-cut has capacity greater than $p-1$, hence at least $p$ by integrality. The same potentials encode a cut of capacity $p$, certifying the exact value without requiring exact flow conservation (see Box~\ref{box:certificate} and Section~\ref{sec:single-certificate}).

\begin{samepage}
\begin{theorem}[Exact max-flow certificate; see Section~\ref{sec:single-certificate}]\label{thm:intro-certificate}

Let $G$ be a directed graph with $n$ vertices, $m$ edges, and integer capacities at most $W\ge1$. For vertices $s,t$ and an integer $p\ge0$, the predicate $\MaxFlow(s,t)=p$ admits a deterministic verifier with a certificate consisting of $n$ vertex potentials, each encoded with $O(\log(nW))$ bits. Some certificate is accepted if and only if $\MaxFlow(s,t)=p$. Verification takes $O(n+m)$ arithmetic operations on $O(\log(nW))$-bit integers.
\end{theorem}
\end{samepage}

Section~\ref{sec:results} and Table~\ref{tab:variants} give the applications of this certificate and their verification bounds.

\begingroup
\widowpenalty=10000
\clubpenalty=10000
\subsection{Results}\label{sec:results}

\endgroup

\begin{table}[H]
\centering
\normalsize
\begin{tabular}{@{}>{\raggedright\arraybackslash}p{0.31\textwidth}>{\raggedright\arraybackslash}p{0.31\textwidth}>{\raggedright\arraybackslash}p{0.31\textwidth}@{}}
\toprule
\textbf{Model}\newline open problem & \textbf{Previous bound} & \textbf{Here}\\
\midrule
\textbf{Online annotated streaming}\newline max-flow value (\cite{Tha16})
  & $\boldsymbol{\tO(m)}$ help and $\boldsymbol{\tO(1)}$ space~\cite{CMT13};\newline $\boldsymbol{\tO(n^2)}$ worst-case total cost
  & $\boldsymbol{\widetilde{O}(n)}$ (help plus space); optimal for online schemes up to logarithmic factors\\
\addlinespace[12pt]
\textbf{Broadcast proof-labeling}\newline max-flow value (\cite{KKP10})
  & {\scriptsize\boldmath$\begin{aligned}[t] &O\bigl(\min\{\alpha(G),p\}\cdot{}\\
    &(\log(p+1)+\log(\Delta(G)+1))\bigr)\end{aligned}$}\newline message bits~\cite{PP22}
  & $\boldsymbol{O(\log n+\log(p+1))}$ label and message bits;\newline optimal for $p\ge n^{\Omega(1)}$\\
\addlinespace[12pt]
\textbf{Merlin--Arthur}\newline all-pairs max-flow (\cite{Akm24})
  & Unit capacities: $\boldsymbol{n^{3.5+o(1)}}$~\cite{Tra25};\newline general: $\boldsymbol{n^{4+o(1)}}$, via deterministic verification
  & $\boldsymbol{\tO(n^3)}$ time; $\boldsymbol{O(n^3\log n)}$ proof bits; independent of $\omega$\\
\addlinespace[12pt]
\textbf{Deterministic verification}\newline all-pairs max-flow (general graphs. \cite{Tra25})
  & $\boldsymbol{n^{4+o(1)}}$ time
  & $\boldsymbol{n^{3.5+o(1)}}$ if $\omega=2$;\newline $\boldsymbol{O(n^{3.614})}$\\
\bottomrule
\end{tabular}
\caption{Verification bounds in the four models. Streaming and all-pairs
bounds are stated for directed graphs with polynomially bounded integer capacities.
Broadcast allows arbitrary integer capacities. Here $p$ is the claimed
max-flow value, $\alpha(G)$ is the arboricity, and $\Delta(G)$ is the maximum
degree. The previous broadcast bound applies to undirected graphs; ours
covers directed graphs too.}
\label{tab:variants}
\end{table}

\begin{samepage}
\paragraph{Streaming verification.}
In the online annotated-streaming model~\cite{CCM09,CCMT14}, the verifier checks a proof against sketches computed during the stream using private randomness. In the model of~\cite{DFPR25}, the prover supplies a certificate before the verifier begins reading the edge stream. This certificate may depend on the whole graph, but not on the stream order, and remains available in read-only memory throughout the stream. Our verifier checks it deterministically in one pass.
\par
\end{samepage}

\begin{theorem}[Streaming maximum-flow values; see Theorem~\ref{thm:streaming} and Corollary~\ref{cor:cert-first}]\label{thm:intro-streaming}

For a directed graph with $n$ vertices, the exact value of $\MaxFlow(s,t)$ admits a one-pass online annotation scheme with polynomially small soundness error, using $\tO(n)$ help and $\tO(n)$ bits of working space. In the certificate-first model, where the certificate is available in read-only memory before the stream, exact-value certification admits a deterministic one-pass verifier with an $\tO(n)$-bit certificate and $\tO(n)$ bits of working space.
\end{theorem}

\begingroup\emergencystretch=2em
Thaler~\cite{Tha16} asks whether computing the maximum-flow value admits an annotation scheme of $o(n^2)$ total cost, which counts both the proof length and the verifier's working space. Theorem~\ref{thm:intro-streaming} achieves $\widetilde O(n)$. The
online total cost and help--space product are optimal up to logarithmic
factors (Corollary~\ref{cor:streaming-optimality}). The certificate-first total cost is also optimal up to logarithmic factors. Section~\ref{sec:streaming} describes the help--space tradeoff.

An earlier scheme for exact maximum-flow values uses $\tO(m)$ online help
and $\tO(1)$ space~\cite{CMT13}.
For bipartite perfect matching, the tradeoff of $\tO(n^2/\ell)$ help and
$\tO(\ell)$ space was already known~\cite{CCMT14}; Chakrabarti, Ghosh,
and Thaler~\cite{CGT20} obtain the same tradeoff for
maximum-matching cardinality in general graphs. We obtain the same tradeoff for exact directed maximum-flow values (Remark~\ref{rem:streaming-tradeoff}). For directed graphs with polynomially bounded integer capacities, Gholizadeh and Jiang~\cite{GJ25} and van den Brand, Song, and Weng~\cite{vdBSW26} give randomized two-party communication protocols using $\widetilde O(n^{3/2})$ bits to compute maximum flows when the edges are partitioned between the parties. The same holds for minimum-cost flows with polynomially bounded integer costs. These protocols compute flows without an untrusted prover.\par\endgroup

\paragraph{Broadcast proof-labeling.}
In distributed verification, each vertex checks a certificate using its own
input and its neighbors' certificates.

\begin{theorem}[Broadcast proof-labeling; see Corollary~\ref{cor:local-max}]\label{thm:intro-labeling}

For directed and undirected graphs with arbitrary nonnegative integer capacities and every integer $p\ge0$, the predicate $\MaxFlow(s,t)=p$ admits a one-round broadcast proof-labeling scheme with $O(\log n+\log(p+1))$ bits per vertex.
\end{theorem}

\begingroup
\clubpenalties 2 10000 0
\widowpenalties 2 10000 0
Korman, Kutten, and Peleg~\cite{KKP10} left the question of optimal bounds for flow verification open.
Their scheme for verifying that the maximum-flow value is $p$ uses
$O(p(\log p+\log n))$ bits per vertex.
For $p\ge n^{\Omega(1)}$, Theorem~\ref{thm:intro-labeling} gives optimal
bounds for this problem in their broadcast model.
Section~\ref{sec:local-max} gives a simple matching lower bound and compares our value-only certificates with schemes that verify a supplied flow or certify only an upper bound on the maximum-flow value.
\par\endgroup

\begingroup
\widowpenalty=10000
\clubpenalty=10000
\paragraph{Merlin--Arthur verification.}
A one-message Merlin--Arthur protocol verifies the exact All-Pairs Max-Flow matrix in $\tO(n^3)$ time.
Merlin sends one proof, after which Arthur uses randomness to accept or reject the claimed matrix.
\par
\endgroup
\begin{theorem}[Cubic Merlin--Arthur verification; see Theorem~\ref{thm:ma}]\label{thm:intro-ma}

For every fixed $c>0$, directed All-Pairs Max-Flow admits a one-message Merlin--Arthur protocol with perfect completeness and soundness error at most $n^{-c}$. The protocol uses $O_c(n^3\log n)$ proof bits, $\tO(n^3)$ verification time, and $\tO(n)$ bits of working space.
\end{theorem}
Akmal~\cite{Akm24} asks: ``Does APC admit a faster
randomized verifier?'' Here APC is directed All-Pairs Connectivity, the
unit-capacity case. The benchmark cited there is a deterministic verifier
taking $O(n^{3.251}+n^{2.5}\sqrt m)$ time, or $O(n^{3.5})$ on dense
graphs; the near-quadratic Merlin--Arthur verifier for
4-Clique~\cite{ACJRW22} motivates the question.
Theorem~\ref{thm:intro-ma} improves the cited dense-graph benchmark
to $\widetilde O(n^3)$ verification time using a Merlin--Arthur protocol,
independently of $\omega$, and extends the result to polynomially bounded integer capacities.
Near-quadratic Merlin--Arthur verification is known for 4-Clique and
All-Pairs Shortest Paths~\cite{ACJRW22}. Whether such a bound is possible
for All-Pairs Connectivity remains open.

\paragraph{Deterministic all-pairs verification.}
Batching the certificates of Theorem~\ref{thm:intro-certificate} gives a subquartic
deterministic verifier for All-Pairs Max-Flow on directed graphs.
The fastest known general algorithm runs an almost-linear-time
maximum-flow algorithm~\cite{CKL+22} separately for every
ordered pair, taking $n^{4+o(1)}$ time on dense graphs.

\begin{theorem}[Subquartic all-pairs verification; see Corollary~\ref{cor:global-all-pairs}]\label{thm:intro-main}

Directed All-Pairs Max-Flow with polynomially bounded integer capacities admits a nondeterministic algorithm with deterministic verification. Its running time is $n^{3.5+o(1)}$ if $\omega=2$, and $O(n^{3.614})$ using published matrix multiplication bounds.
\end{theorem}

\begingroup
Theorem~\ref{thm:intro-main} resolves the case of polynomially bounded integer capacities beyond unit capacities in Trabelsi's question~\cite{Tra25} about subquartic nondeterministic algorithms for directed All-Pairs Max-Flow.
The bound in Theorem~\ref{thm:intro-main} uses published rectangular
multiplication bounds~\cite{WXXZ24}.
For all-pairs verification, supplied cut capacities can already be checked
deterministically in $n^{\omega(2,1,1)+o(1)}$ time~\cite{Tra25}.
The remaining challenge is to verify the flow lower bounds. We
nevertheless encode a cut in the potentials, so that each certificate establishes both bounds.
\par
\endgroup

\begin{remark}[Fine-grained implications]\label{rem:fine-grained}
Our verification bounds are unconditional. The non-reducibility argument
of~\cite{Tra25} also applies to deterministic fine-grained reductions
from unrestricted CNF-SAT under the following hypothesis: for every
constant $\varepsilon>0$, CNF-UNSAT on $N$ variables and input length
$L$ has no nondeterministic $2^{(1-\varepsilon)N}L^{O(1)}$-time
algorithm. We call this unrestricted-width hypothesis NSETH$'$, by
analogy with SETH$'$~\cite{Bri14} and the nondeterministic strong
exponential time hypothesis (NSETH)~\cite{CGIMPS16}.
The recently announced subexponential-time bounds for fixed-width SAT~\cite{OAI26eth,Ant26eth}
do not by themselves refute NSETH$'$: their guarantees are for each fixed
clause width $k$, with width-dependent bounds, whereas NSETH$'$ concerns
CNF formulas of unrestricted width.
\end{remark}

\subsection{Technical overview}\label{sec:overview}

\paragraph{The vertex check.}
For a potential vector $y$, write
\begin{equation*}\tag{$\star$}\label{eq:star}
  \operatorname{ex}_y(v)\;=\;\sum_{u} c_{vu}\operatorname{clip}(y_v-y_u,0,1)
  \;-\;\sum_{u} c_{uv}\operatorname{clip}(y_u-y_v,0,1).
\end{equation*}
As described in Box~\ref{box:certificate}, we check this expression against its target excess to within $1/(8n)$. Each incident edge's contribution depends only on its capacity and the potentials at its two endpoints. The check therefore gives broadcast verification directly, and deterministic streaming verification when the potentials precede the stream. For broadcast verification, capping capacities at $p+1$ makes the required precision depend on $p$ rather than $W$.

The classical clipped-potential representation has three properties used by our verification algorithms:
\begin{enumerate}[label=(\roman*),ref=(\roman*),leftmargin=*]
\item\label{item:capacity-separation} capacities factor out of the potential-dependent terms,
\item\label{item:common-thresholds} all edges share the same clipping thresholds, and
\item\label{item:linear-middle} the middle range is linear.
\end{enumerate}
For the energy $\sum_{e:c_e>0}f_e^2/(2c_e)$, the derivative with respect to $f_e$ is $f_e/c_e$; classical optimality conditions~\cite{Roc84,AMO93} give $f_e=c_e\clip(y_u-y_v,0,1)$. Capacity separation~\ref{item:capacity-separation} lets annotated streaming and MA verification summarize the graph before reading the potentials. The common thresholds~\ref{item:common-thresholds} allow one sorted order per terminal pair, and linearity~\ref{item:linear-middle} reduces whole ranges to two weighted sums for fast MA evaluation and deterministic batching. By contrast, unweighted quadratic energy would give $f_e=\clip(y_u-y_v,0,c_e)$, with capacity-dependent saturation thresholds that complicate these algorithms.

\paragraph{Checking excesses through capacity sketches.}
The verifier must summarize the capacities before it knows the clipped
factors in~\eqref{eq:star}. We encode the capacities and the clipped factors
separately as polynomials. Given $n$ values indexed by the vertices
$1,\ldots,n$, \textit{interpolation} gives a polynomial of degree at most $n-1$
that takes the prescribed value at each vertex index. For each $j$, we
use this to define polynomials satisfying
\[
\begin{aligned}
\mathcal A_j^+(v)&=c_{vj},
&\mathcal A_j^-(v)&=c_{jv},\\
\mathcal K_j^+(v)&=Q_0\operatorname{clip}(y_v-y_j,0,1),\\
\mathcal K_j^-(v)&=Q_0\operatorname{clip}(y_j-y_v,0,1)
\end{aligned}
\]
at every vertex index $v$. Here $Q_0$ is the grid denominator from
Box~\ref{box:certificate}, so the scaled clipped factors are integers.
Importantly, the $\mathcal A$-polynomials depend only on the graph, while
the $\mathcal K$-polynomials depend only on the supplied potentials.
Over a sufficiently large finite field, the polynomial
\[
\mathcal E(X)=\sum_{j=1}^n
 \bigl(\mathcal A_j^+(X)\mathcal K_j^+(X)
      -\mathcal A_j^-(X)\mathcal K_j^-(X)\bigr)
\]
has degree at most $2n-2$. By the definitions above, $\mathcal E(v)=Q_0\operatorname{ex}_y(v)$ in the field at every vertex index $v$.

The verifier chooses a private point $\zeta$. At this point, each
interpolating polynomial is a weighted sum of its prescribed values,
with weights $\lambda_1,\ldots,\lambda_n$ determined by $\zeta$ and the
vertex indices. Thus
\[
\mathcal A_j^+(\zeta)=\sum_u\lambda_u c_{uj},
\qquad
\mathcal A_j^-(\zeta)=\sum_u\lambda_u c_{ju}.
\]
The verifier computes the weights before the stream and maintains these
$2n$ sums. An arriving edge $(u,v)$ of capacity $c_{uv}$ adds
$\lambda_uc_{uv}$ to $\mathcal A_v^+(\zeta)$ and $\lambda_vc_{uv}$ to
$\mathcal A_u^-(\zeta)$. Capacity updates are handled by adding the
corresponding weighted change.
Once the potentials arrive, their clipped factors determine the values
$\mathcal K_j^\pm(\zeta)$, and the displayed identity gives
$\mathcal E(\zeta)$. The prover supplies $2n-1$ claimed evaluations at
$1,\ldots,2n-1$, whose first $n$ entries claim the scaled vertex excesses.
These determine a polynomial $H$ of degree at most $2n-2$. The verifier
checks the claimed excesses against their targets and tests
$H(\zeta)=\mathcal E(\zeta)$.

These computations are modulo a prime $P$, so different integers can
appear equal if their difference is a multiple of $P$. The capacities
bound the magnitude of every true scaled excess, and the verifier checks
the claimed scaled excesses against the same bound. We choose $P$ larger
than twice this bound, so two allowed excesses agree modulo $P$ only when
they are equal as integers. An incorrect claimed excess therefore gives
$H(v)\ne\mathcal E(v)$ at some vertex, making $H-\mathcal E$ a nonzero
polynomial. Since the
proof is fixed independently of $\zeta$, the polynomial root bound gives
soundness.

Thaler's Property-B check~\cite{Tha16} uses a polynomial to count edges violating claimed component labels. These counts are nonnegative, so checking that their sum is zero rules out violations. Our polynomial instead encodes signed scaled pseudoflow excesses, which sum to zero even when the required balance constraints fail. We therefore compare each vertex's excess with its target, allowing the small error introduced by rounding. Together with the encoded cut, these checks certify the exact maximum-flow value by integrality.

The shared setup in Section~\ref{sec:applications} gives the encoding
and field choices.
The capacity sketches depend only on the graph and the private point,
so the same sketches serve every terminal pair. This is the common
algebraic step behind online annotated streaming and Merlin--Arthur
verification. For MA, it remains to evaluate the clipped factors quickly
for each supplied potential vector; the next idea does this without
reconstructing every edge amount.

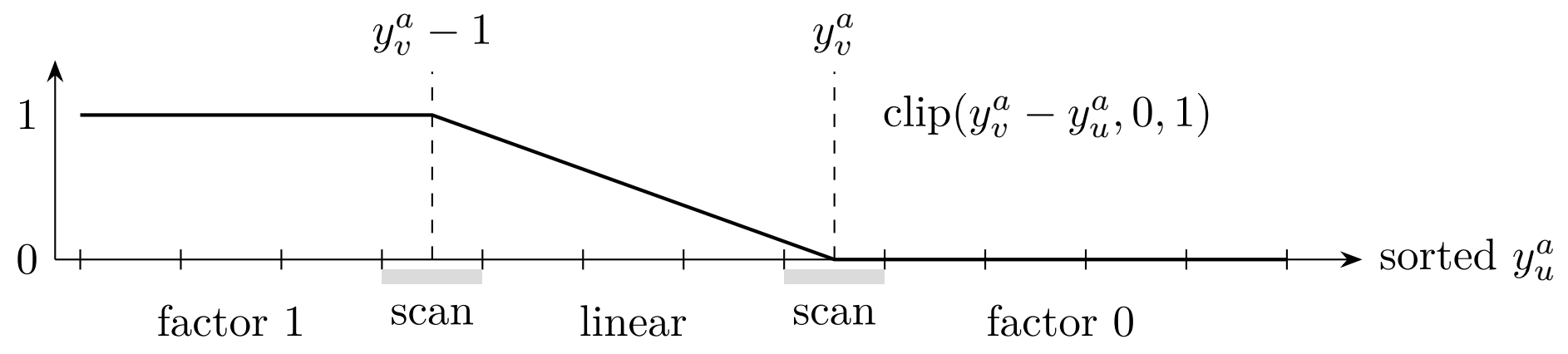
\begin{figure}[H]
\centering
\begin{tikzpicture}[x=1cm,y=1.15cm,font=\small]
  % axes
  \draw[-{Stealth[length=5pt]}] (0,0) -- (10.4,0)
    node[anchor=west] {sorted $y^a_u$};
  \draw[-{Stealth[length=5pt]}] (0,0) -- (0,1.38);
  \node[anchor=east] at (0,0) {$0$};
  \node[anchor=east] at (0,1) {$1$};
  % breakpoints
  \draw[dashed] (3.0,0) -- (3.0,1.3) node[above] {$y^a_v-1$};
  \draw[dashed] (6.2,0) -- (6.2,1.3) node[above] {$y^a_v$};
  % the clipped factor
  \draw[thick] (0.2,1) -- (3.0,1) -- (6.2,0) -- (9.8,0);
  \node[anchor=west] at (6.45,1.0) {$\operatorname{clip}(y^a_v-y^a_u,0,1)$};
  % buckets of B consecutive vertices
  \foreach \x in {0.2,1.0,...,9.81} \draw (\x,-0.07) -- (\x,0.07);
  \fill[gray!28] (2.6,-0.17) rectangle (3.4,-0.07);
  \fill[gray!28] (5.8,-0.17) rectangle (6.6,-0.07);
  \node[anchor=north] at (1.4,-0.2) {factor $1$};
  \node[anchor=north] at (3.0,-0.2) {scan};
  \node[anchor=north] at (4.6,-0.2) {linear};
  \node[anchor=north] at (6.2,-0.2) {scan};
  \node[anchor=north] at (8.0,-0.2) {factor $0$};
\end{tikzpicture}
\caption{Clipping after sorting. Whole buckets within one of the three
regimes use precomputed capacity and potential-weighted capacity sums. Only the breakpoint
buckets (shaded) are scanned, at most two per orientation. The incoming
check has breakpoints $y^a_v$ and $y^a_v+1$.}\label{fig:blocking}
\end{figure}

\paragraph{Evaluating clipped sums after sorting.}
Fix one terminal pair and let $y$ be its potential vector. Sort the vertices
$u$ by increasing $y_u$. For a fixed vertex $v$, the factors
$\operatorname{clip}(y_v-y_u,0,1)$ then form three consecutive ranges in
this order: saturated when $y_u\le y_v-1$, linear when
$y_v-1<y_u<y_v$, and zero when $y_u\ge y_v$
(Figure~\ref{fig:blocking}). We reuse the same sorted order when checking
every vertex $v$. For the
MA protocol, the coefficients multiplying these factors are the
interpolation weights $\lambda_u$. On a linear interval $I$, the
contribution is
\[
\sum_{u\in I}\lambda_u(y_v-y_u)
=y_v\sum_{u\in I}\lambda_u-\sum_{u\in I}\lambda_uy_u.
\]
We compute prefix sums of $\lambda_u$ and $\lambda_u y_u$ along the sorted order. If an interval $I$ occupies positions $\ell,\ldots,r$, each required sum is the corresponding prefix sum through position $r$ minus the prefix sum through position $\ell-1$. The
saturated range contributes only the sum of its $\lambda_u$'s, and the
zero range contributes nothing. The range boundaries occur at $y_v-1$
and $y_v$. As $v$ moves through the sorted order, both boundaries move
only forward, so two pointers locate all of them in one sweep. The
reverse clipped factors are handled similarly. Multiplying by $Q_0$
gives the scaled values $\mathcal K_j^\pm(\zeta)$, completing the fast
evaluation for each pair in the MA protocol (Theorem~\ref{thm:ma} and
Lemma~\ref{lem:ma-clipped}).

Deterministic verification uses the same three ranges, but must evaluate
the actual vertex excesses for all pairs. Checking each of the $n(n-1)$
certificates separately against a dense graph would take $\Theta(n^4)$
time. For each terminal pair, we divide its sorted vertex order into buckets of
$B$ consecutive vertices. The order and bucket partition depend on the
terminal pair and are reused for every vertex $v$ being checked.

A whole bucket $I$ lies in one range if all its vertices $u$ belong to the
same clipping regime relative to the thresholds $y_v-1$ and $y_v$. If the
bucket lies wholly in the linear range, its outgoing contribution is
\[
\sum_{u\in I}c_{vu}(y_v-y_u)
=y_v\sum_{u\in I}c_{vu}-\sum_{u\in I}y_uc_{vu}.
\]
Thus we need only the total capacity and the potential-weighted capacity
of the bucket. A saturated bucket contributes its total capacity, and a
zero bucket contributes nothing. Incoming contributions use the analogous
two sums with capacities $c_{uv}$, with thresholds $y_v$ and $y_v+1$. Here the coefficients are capacities
$c_{vu}$ or $c_{uv}$, which vary with the vertex being checked. We compute
their different bucket sums together using matrix multiplication. Let
$\mathbf A=(c_{uv})$ be the capacity matrix. For each terminal pair $a$
and each of its buckets $I$, create one row of $\mathbf X$ containing $1$
on $I$ and $0$ elsewhere. The corresponding row of $\mathbf Z$ contains
the integer potential indices $Q_0y_u^a$ on $I$ and $0$ elsewhere. The
four products
\[
\mathbf X\mathbf A,\qquad
\mathbf X\mathbf A^\top,\qquad
\mathbf Z\mathbf A,\qquad
\mathbf Z\mathbf A^\top
\]
give the incoming capacities, outgoing capacities, and the same two sums
weighted by the potential indices, respectively. Dividing the last two
outputs by $Q_0$ gives the potential-weighted capacity sums. Thus four
products provide the required aggregates for every pair, bucket, and
vertex together.

Prefix sums of the bucket aggregates give the sums over any run of whole
buckets by subtraction. Only buckets containing a clipping breakpoint
need to be scanned individually: at most two for the outgoing sum and
two for the incoming sum. Larger buckets give fewer rows in the matrix
products, but leave more vertices to scan around each breakpoint.
Choosing $B$ to balance these costs gives the deterministic bound of
Corollary~\ref{cor:global-all-pairs}.

Given a set of $n$ vectors in $\mathbb R^n$, Matou\v{s}ek's dominance-product algorithm~\cite{Mat91} counts, for each pair $x,z$ from this set, the coordinates $i$ satisfying $x_i\le z_i$. He sorts the entries in each coordinate into buckets, uses matrix multiplication to count comparisons between different buckets, and checks comparisons within a bucket directly. A barrier in adapting this idea is that dominance comparisons depend only on the relative order of coordinate values, while our clipped flow values depend on actual potential differences and edge capacities. Our matrix products therefore incorporate capacities and potential values, rather than only the bucket membership information needed to count comparisons. The numerical bound and exact-arithmetic implementation are given in the proof of Corollary~\ref{cor:global-all-pairs}.

Section~\ref{sec:single-certificate} proves the certificate theorem. Section~\ref{sec:applications} develops the applications, and Section~\ref{sec:open} lists open problems. The examples below illustrate the certificate and its motivation.

\subsection{Large denominators and dense flow witnesses}\label{sec:illustrations}
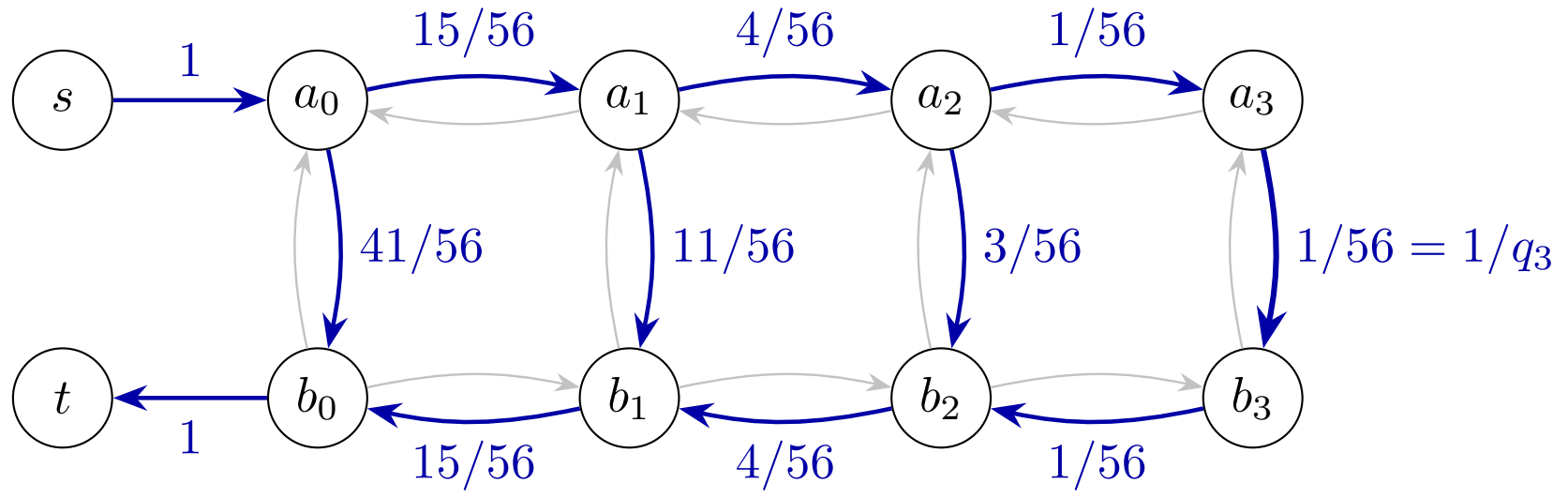
\begin{figure}[htbp]
\centering
\begin{tikzpicture}[x=1cm,y=1cm,>=Stealth,font=\small,
  vertex/.style={circle,draw,fill=white,minimum size=7mm,inner sep=1pt},
  positive/.style={->,blue!65!black,line width=.85pt},
  reverse/.style={->,gray!48,line width=.45pt},
  amount/.style={fill=white,inner sep=1.4pt,font=\small,text=blue!65!black}]
\foreach \i/\x in {0/0,1/2.2,2/4.4,3/6.6}{
  \node[vertex] (a\i) at (\x,2.1) {$a_{\i}$};
  \node[vertex] (b\i) at (\x,0) {$b_{\i}$};
}
\node[vertex] (s) at (-1.8,2.1) {$s$};
\node[vertex] (t) at (-1.8,0) {$t$};
% Zero-flow reverse arcs are drawn first.
\foreach \i/\j in {0/1,1/2,2/3}{
  \draw[reverse] (a\j) to[bend left=12] (a\i);
  \draw[reverse] (b\i) to[bend left=12] (b\j);
}
\foreach \i in {0,1,2,3}{
  \draw[reverse] (b\i) to[bend left=12] (a\i);
}
% The positive-flow arcs form the electrical flow on the ladder.
\draw[positive] (s) -- node[amount,above=3pt] {$1$} (a0);
\draw[positive] (b0) -- node[amount,below=3pt] {$1$} (t);
\foreach \i/\j/\f in {0/1/15,1/2/4,2/3/1}{
  \draw[positive] (a\i) to[bend left=12]
    node[amount,above=2pt] {$\f/56$} (a\j);
  \draw[positive] (b\j) to[bend left=12]
    node[amount,below=2pt] {$\f/56$} (b\i);
}
\foreach \i/\f in {0/41,1/11,2/3}{
  \draw[positive] (a\i) to[bend left=12]
    node[amount,right=2pt] {$\f/56$} (b\i);
}
\draw[positive,line width=1.15pt] (a3) to[bend left=12]
  node[amount,right=2pt] {$1/56=1/q_3$} (b3);
\end{tikzpicture}
\caption{The unit-capacity ladder described in Subsection~\ref{sec:illustrations}, with $k=3$.
Blue labels give the minimum-energy flow amounts on blue arcs; gray
reverse arcs carry zero. Every arc has capacity one. The last rung
carries $1/56=1/q_3$, illustrating the growth of the exact denominator.}
\label{fig:ladder-denominator}
\end{figure}

Even unit capacities can give exponentially large denominators in the
minimum-energy flow. Any capacity-feasible clipped-potential assignment
with a prescribed excess vector minimizes the quadratic energy for that
vector: on each edge, the clipped formula gives the corresponding
first-order optimality condition, and these conditions are sufficient by
convexity. Strict convexity makes the minimizing edge assignment unique.
Thus an exactly conserving clipped-potential representation must
reproduce the minimum-energy flow. Take ladder vertices $a_0,\ldots,a_k$ and
$b_0,\ldots,b_k$, with both directions of every rail edge
$a_i a_{i+1}$ and $b_i b_{i+1}$ and every rung $a_i b_i$.
Add terminals $s,t$ and arcs $s\to a_0$ and $b_0\to t$.
Every arc has capacity one. The maximum-flow value is one, and both
terminal arcs are full in every maximum flow.
Figure~\ref{fig:ladder-denominator} shows the case $k=3$.

% All arcs, including gray reverse arcs,
% have capacity one. Blue labels are flow amounts, not capacities.

Canceling equal flow on opposite arcs preserves excesses and lowers
energy, so the remaining problem reduces to unit electrical flow on
the undirected ladder. By symmetry, write its top and bottom potentials
as $d_i/2$ and $-d_i/2$. For $k\ge1$, conservation gives
\[
\begin{aligned}
3d_0-d_1&=2,\\
4d_i-d_{i-1}-d_{i+1}&=0\quad(1\le i<k),\\
3d_k-d_{k-1}&=0.
\end{aligned}
\]
The rung amounts are $d_i$, and each rail carries
$(d_i-d_{i+1})/2$, toward increasing indices on the top rail and toward
decreasing indices on the bottom rail. The solution has
$0<d_k<\cdots<d_0<1$, so these amounts respect all unit capacities.
Assigning each amount to the corresponding directed arc and zero to its
reverse therefore gives the unique minimum-energy maximum flow in the
directed network.

To verify the solution and its denominator, set $q_{-1}=0$, $q_0=1$, and $q_j=4q_{j-1}-q_{j-2}$ for $j\ge1$. Direct substitution into the conservation equations gives
\[
d_i=\frac{q_{k-i}-q_{k-i-1}}{q_k}\quad(0\le i\le k),
\qquad d_k=\frac1{q_k}.
\]
The differences $q_j-q_{j-1}$ are positive and strictly increasing for $j\ge0$, and $d_0=1-q_{k-1}/q_k<1$ for $k\ge1$. This also verifies the claimed inequalities $0<d_k<\cdots<d_0<1$. The denominators are $q_0,q_1,q_2,\ldots=1,4,15,56,209,\ldots$. The integer $q_k$ is also the number of spanning trees of the underlying undirected ladder, which has $k+1$ rungs~\cite{SHB09}. The recurrence gives $q_j\ge3q_{j-1}$ for $j\ge1$, and hence $q_k\ge3^k$, while the graph has $2k+4$ vertices.
Thus the last rung's flow has an exponentially large denominator even
in lowest terms. Rounding the potentials avoids these exact denominators.

The following example shows a maximum flow and its potential certificate.

\begin{example}\label{ex:small}
Figure~\ref{fig:small-flow} shows a six-vertex network with a flow of value $6$. Edge labels give \emph{flow/capacity}: the outer edges are full, while every blue middle edge uses half its capacity. Each intermediate vertex receives and sends $3=2+1$ units. The cut $\{s\}$ has capacity $6$, so this flow is maximum.

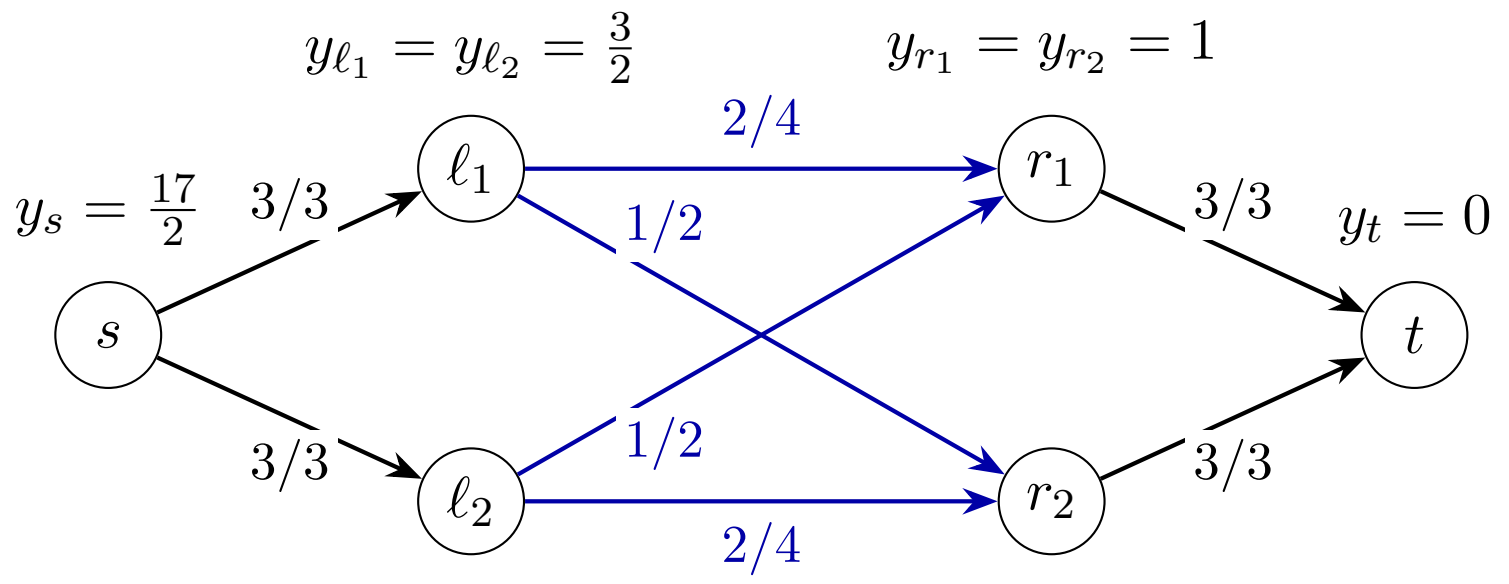
\begin{figure}[H]
\centering
\begin{tikzpicture}[x=1.2cm,y=1cm,>=Stealth,
vertex/.style={circle,draw,fill=white,minimum size=7mm,inner sep=1pt},
amount/.style={fill=white,inner sep=1.5pt,font=\small}]
\node[vertex] (s) at (0,0) {$s$};
\node[vertex] (l1) at (2,1.1) {$\ell_1$};
\node[vertex] (l2) at (2,-1.1) {$\ell_2$};
\node[vertex] (r1) at (5.2,1.1) {$r_1$};
\node[vertex] (r2) at (5.2,-1.1) {$r_2$};
\node[vertex] (t) at (7.2,0) {$t$};
\draw[->,thick] (s)--node[amount,above=2pt] {$3/3$} (l1);
\draw[->,thick] (s)--node[amount,below=2pt] {$3/3$} (l2);
\draw[->,thick] (r1)--node[amount,above=2pt] {$3/3$} (t);
\draw[->,thick] (r2)--node[amount,below=2pt] {$3/3$} (t);
\draw[->,thick,blue!65!black] (l1)--node[amount,above=2pt] {$2/4$} (r1);
\draw[->,thick,blue!65!black] (l2)--node[amount,below=2pt] {$2/4$} (r2);
\draw[->,thick,blue!65!black] (l1)--node[amount,pos=.30,above=3pt] {$1/2$} (r2);
\draw[->,thick,blue!65!black] (l2)--node[amount,pos=.30,below=3pt] {$1/2$} (r1);
\node[above=3pt of s] {$y_s=\tfrac{17}{2}$};
\node at (2,1.90) {$y_{\ell_1}=y_{\ell_2}=\tfrac32$};
\node at (5.2,1.90) {$y_{r_1}=y_{r_2}=1$};
\node[above=3pt of t] {$y_t=0$};
\end{tikzpicture}
\caption{A maximum flow and its potential certificate on six vertices.}
\label{fig:small-flow}
\end{figure}

These potentials reproduce the displayed flow through~\eqref{eq:clip}: the outer drops are at least $1$, and every middle drop is $1/2$. The source potential $y_s=17/2$ lies in the upper interval $[6,11]$, while all other potentials lie in $[0,5]$. Thus the encoded cut is $S=\{s\}$.

\end{example}

\subsubsection{Dense maximum flows}\label{sec:dense-proof}
\begin{figure}[htbp]
\centering
\begin{tikzpicture}[x=1cm,y=1cm,>=Stealth,font=\small,
 vertex/.style={circle,draw=black!65,fill=white,minimum size=3.5mm,inner sep=0pt},
 active/.style={->,blue!65!black,line width=.75pt},
 inactive/.style={->,gray!48,line width=.45pt},
 outeractive/.style={->,blue!65!black,line width=1.45pt},
 outerinactive/.style={->,gray!48,line width=.85pt}]
% Four layers; three representative vertices give the k=3 instance.
\foreach \x/\name in {0/S,2.7/L,5.4/R,8.1/T}{
 \filldraw[rounded corners=5pt,fill=gray!5,draw=gray!30]
  (\x-.38,-1.43) rectangle (\x+.38,1.43);
 \node[font=\small\bfseries] at (\x,1.78) {$\name$};
 \node[font=\scriptsize] at (\x,-1.72) {$k$ vertices};
}
\foreach \i/\y in {1/1.05,2/0,3/-1.05}{
 \coordinate (S\i) at (0,\y);
 \coordinate (L\i) at (2.7,\y);
 \coordinate (R\i) at (5.4,\y);
 \coordinate (T\i) at (8.1,\y);
}
% Zero-flow edges for the selected terminal pair.
\foreach \i in {2,3}{\foreach \j in {1,2,3}{
 \draw[outerinactive,shorten <=2mm,shorten >=2mm] (S\i)--(L\j);
}}
\foreach \i in {1,2,3}{\foreach \j in {1,2}{
 \draw[outerinactive,shorten <=2mm,shorten >=2mm] (R\i)--(T\j);
}}
% A value-k^2 maximum flow: k units on active outer edges, 1 in the middle.
\foreach \j in {1,2,3}{
 \draw[outeractive,shorten <=2mm,shorten >=2mm] (S1)--(L\j);
 \draw[outeractive,shorten <=2mm,shorten >=2mm] (R\j)--(T3);
}
\foreach \i in {1,2,3}{\foreach \j in {1,2,3}{
 \draw[active,shorten <=2mm,shorten >=2mm] (L\i)--(R\j);
}}
\foreach \i in {1,2,3}{
 \node[vertex] at (S\i) {};
 \node[vertex] at (L\i) {};
 \node[vertex] at (R\i) {};
 \node[vertex] at (T\i) {};
}
\node[vertex,draw=blue!65!black,line width=1pt] at (S1) {};
\node[vertex,draw=blue!65!black,line width=1pt] at (T3) {};
\node[anchor=east,text=blue!65!black] at (-.48,1.05) {$s$};
\node[anchor=west,text=blue!65!black] at (8.58,-1.05) {$t$};
\node at (1.35,1.78) {capacity $k$};
\node[text=blue!65!black] at (4.05,1.78) {capacity $1$};
\node at (6.75,1.78) {capacity $k$};
\end{tikzpicture}
\caption{The four-layer graph of Observation~\ref{obs:dense}, illustrated for $k=3$. Each consecutive pair of layers is complete bipartite, with capacities $k,1,k$ from left to right. For the highlighted terminals $s,t$, blue outer edges carry $k$ units each, every blue middle edge carries one unit, and gray edges carry zero. Every choice of $s\in S,t\in T$ forces all $k^2$ middle edges to be saturated.}
\label{fig:dense-construction}
\end{figure}
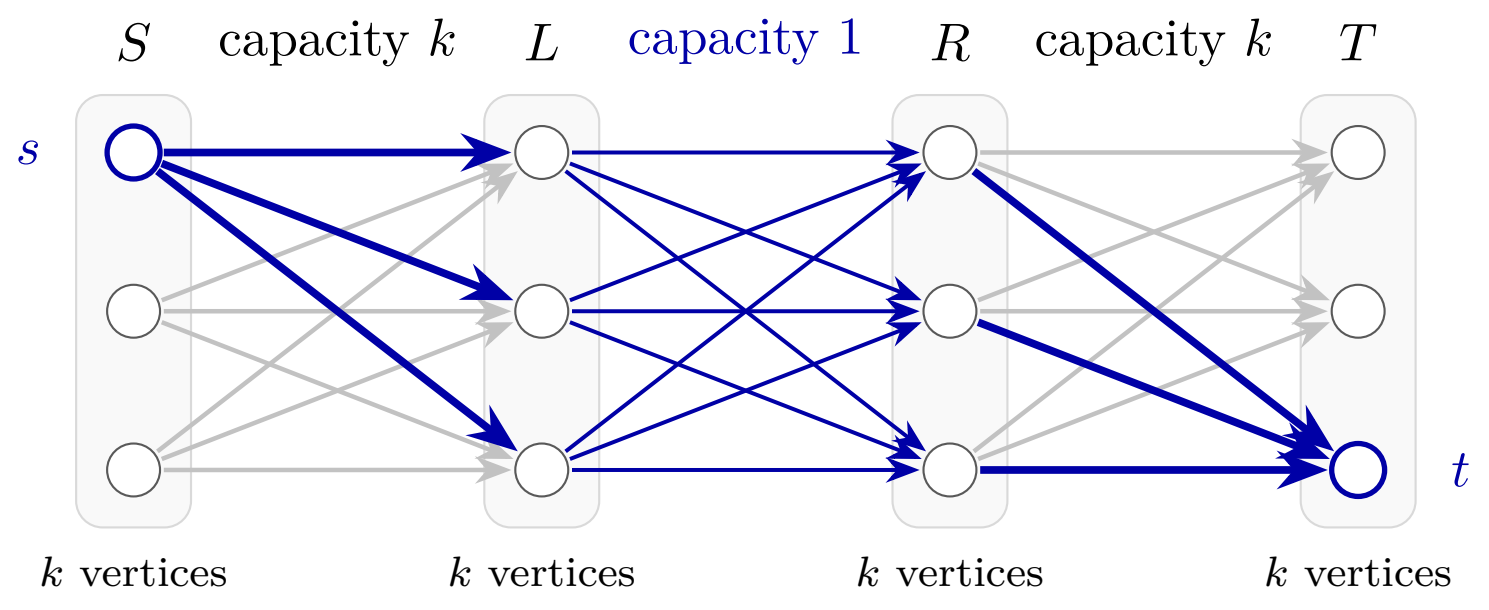

\begin{observation}\label{obs:dense}
For every $n\ge8$ divisible by $4$, there is a directed graph with capacities in $\{1,n/4\}$ such that, for $n^2/16$ ordered pairs, every maximum flow is positive on $\Theta(n^2)$ unit-capacity edges. Consequently, explicitly listing a maximum flow for each of these pairs requires $\Omega(n^4)$ edge amounts in total.
\end{observation}
\begin{proof}[Proof of Observation~\ref{obs:dense}]
Put $k=n/4$ and take four disjoint vertex sets $S,L,R,T$, each of size $k$. Include all edges from $S$ to $L$ and from $R$ to $T$ with capacity $k$, and all edges from $L$ to $R$ with capacity one.
Fix $s\in S$ and $t\in T$. Sending $k$ units from $s$ to every vertex of $L$, one unit on every edge of $L\times R$, and $k$ units from every vertex of $R$ to $t$ gives a feasible flow of value $k^2$. The cut $S\cup L$ has exactly $k^2$ outgoing edges, all of capacity one, and no incoming edges. Thus the maximum-flow value is $k^2$, and every maximum flow must saturate all $k^2$ edges of $L\times R$.
There are $|S||T|=k^2=n^2/16$ choices of $(s,t)$, each requiring $k^2=\Theta(n^2)$ explicit middle-edge amounts. Hence listing one maximum flow for every such pair requires at least $k^4=\Omega(n^4)$ edge amounts. Figure~\ref{fig:dense-construction} illustrates the construction.
\end{proof}

\section{Preliminaries}\label{sec:prelim}
Unless stated otherwise, we work with a directed graph $G=(V,E)$ on $n=|V|$ vertices and $m=|E|$ edges. We exclude self-loops and allow at most one edge per ordered pair; antiparallel edges are permitted. Every edge $(u,v)$ has a nonnegative integer capacity $c_{uv}$, and we set $c_{uv}=0$ when the edge is absent. We write $W\ge1$ for an upper bound on the capacities and assume $W=n^{O(1)}$ whenever a bound is stated only in terms of $n$, unless specified otherwise. Terminals $s,t$, when present, are distinct. The capacity-weighted out- and in-degrees of a vertex $v$ are $d_c^+(v)=\sum_u c_{vu}$ and $d_c^-(v)=\sum_u c_{uv}$, respectively.

We write $\omega$ for the square matrix multiplication exponent and $\omega(a,b,c)$ for the exponent of multiplying an $n^a\times n^b$ matrix by an $n^b\times n^c$ matrix. Running times stated with these exponents hide $n^{o(1)}$ factors, and $\tO(\cdot)$ also hides polylogarithmic factors.

\begin{definition}[Capacity-feasible edge assignment]
An edge assignment $f$ is capacity-feasible if $0\le f_{uv}\le c_{uv}$ for every ordered pair $(u,v)$.
\end{definition}
\begin{definition}[Signed excess]
We use the outflow-minus-inflow convention:
\begin{equation}\label{eq:excess}
\ex_f(v)=\sum_{u\in V}f_{vu}-\sum_{u\in V}f_{uv}.
\end{equation}
Thus $\ex_f$ is net outflow: positive values mean supply and negative values mean demand. An exact $st$-flow of value $F$ satisfies
\[
\ex_f(s)=F,\qquad\ex_f(t)=-F,\qquad\ex_f(v)=0\quad(v\notin\{s,t\}).
\]
A capacity-feasible edge assignment that may violate these equalities is called a pseudoflow.
\end{definition}
For $S\subseteq V$, define
\[
\delta^+(S)=\{(u,v):u\in S,v\notin S\},\qquad
\delta^-(S)=\{(u,v):u\notin S,v\in S\}.
\]
If $s\in S$ and $t\notin S$, then $S$ is an $st$-cut, with capacity $c(\delta^+(S))=\sum_{(u,v)\in\delta^+(S)}c_{uv}$.
For every edge assignment $f$, summing the excesses over $S$ cancels internal edges, giving
\begin{equation}\label{eq:cut-identity}
\sum_{v\in S}\ex_f(v)=f(\delta^+(S))-f(\delta^-(S)).
\end{equation}
For a capacity-feasible assignment,
\begin{equation}\label{eq:cut-bound}
\sum_{v\in S}\ex_f(v)\le f(\delta^+(S))\le c(\delta^+(S)).
\end{equation}
By the max-flow/min-cut theorem~\cite{FF56},
\begin{equation}\label{eq:maxflow-mincut}
\MaxFlow(s,t)=\min_{S:s\in S,\,t\notin S}c(\delta^+(S)).
\end{equation}

We use the total-function notion of nondeterministic time from~\cite{AKT21}, as restated in~\cite{Tra25}.
\begin{definition}[Nondeterministic algorithm for a total function]
A nondeterministic algorithm for a total function $\mathcal F$ with running time $T(n)$ is a deterministic algorithm $\mathcal V$ that receives an instance $x$ and a guess $g$ and satisfies the following: $\mathcal V(x,g)$ runs in time $T(n)$ for every $g$; $\mathcal V(x,g)$ outputs either a solution for $x$ or ``don't know'' for every $g$; and $\mathcal V(x,g)$ outputs a solution for at least one $g$.
\end{definition}

\paragraph{Polynomial evaluation check.}\label{par:polynomial-check}
Both randomized verifiers use the standard low-degree-extension approach
to annotated streams~\cite{CCMT14}, in the form of Thaler's
Property-B polynomial check \cite{Tha16}.
Let $E$ have degree at most $d$ over a prime field $\mathbb F_P$, with
$P>d+1$. A prover supplies claimed evaluations $h_1,\ldots,h_{d+1}$
at the public points $1,\ldots,d+1$. These specify the unique polynomial
\[
H(X)=\sum_{i=1}^{d+1}h_i\psi_i(X),\qquad \deg H\le d,
\]
where $\psi_i$ is the Lagrange basis on these points. If the verifier can
evaluate the actual $E$ at a private point $\zeta$, it checks
$H(\zeta)=E(\zeta)$. Fix $E$ and the table independently of $\zeta$.
When $H\ne E$, their difference has at most $d$ roots, so the error is
at most $d/P$ for uniform $\zeta$, or $2d/P$ when each point has
probability at most $2/P$. Honest evaluations give $H=E$ and pass for
every point. The same point can check a family of fixed tables without a union
bound. If any table is incorrect, fix one such table. Acceptance
implies that its polynomial check passes, so the error bound for that
single table applies.

The weights $\psi_i(\zeta)$ on consecutive nodes take $\tO(d+1)$ time
to compute, after which evaluating $H(\zeta)$ takes $O(d+1)$ field
operations. Evaluating the actual $E(\zeta)$ is the application-specific step.

\section{The max-flow certificate}\label{sec:single-certificate}
Throughout this section, $G=(V,E)$ is a directed network with $n$ vertices, $m$ edges, and nonnegative integer capacities bounded by an integer $W\ge1$.
Fix $s\ne t$ and a claimed integer $p\ge0$. Put $\varepsilon_0=1/(8n)$ and $b=p(\evec_s-\evec_t)$. Supply one potential per vertex, subject to
\begin{equation}\label{eq:cut-encoding}
\begin{gathered}
y_v\in[0,n-1]\cup[n,2n-1]\quad(v\in V),\\
y_s\ge n,\qquad y_t\le n-1.
\end{gathered}
\end{equation}
The potentials define $f$ by~\eqref{eq:clip}, and the verifier checks
\begin{equation}\label{eq:approx-ex}
|\ex_f(v)-b(v)|\le\varepsilon_0\qquad\text{for every }v\in V.
\end{equation}
The vertices with potentials in the upper interval form an $st$-cut, letting one vector certify both bounds. With adjacency-list input, the verifier checks~\eqref{eq:cut-encoding} and~\eqref{eq:approx-ex} in $O(n+m)$ word operations on $O(\log(nW))$-bit integers, where $m$ is the number of edges.

\subsection{Soundness (approximate conservation is enough)}
\begin{lemma}[A lower bound from approximate conservation]\label{cor:mf-lower}
Let $N\ge n$ and $p\ge0$ be integers. If a capacity-feasible assignment $g$
satisfies $|\ex_g(v)-p(\one[v=s]-\one[v=t])|\le1/(8N)$ at every vertex,
then $\MaxFlow(s,t)\ge p$.
\end{lemma}
\begin{proof}
Fix an $st$-cut $A$. Its target excesses sum to $p$, since it contains $s$ but not $t$. Summing the vertex errors and using~\eqref{eq:cut-bound} gives
\[
c(\delta^+(A))\ge \sum_{v\in A}\ex_g(v)
\ge p-\frac{|A|}{8N}>p-1.
\]
Both $c(\delta^+(A))$ and $p$ are integers, so this cut has capacity at least $p$. The argument holds for every $st$-cut; the max-flow/min-cut theorem~\eqref{eq:maxflow-mincut} therefore gives $\MaxFlow(s,t)\ge p$.
\end{proof}
\begin{lemma}[Single-pair correctness]\label{cor:encoded-value}
If the interval conditions~\eqref{eq:cut-encoding} and excess checks~\eqref{eq:approx-ex} pass, then $\MaxFlow(s,t)=p$.
\end{lemma}
\begin{proof}
Let $S=\{v:y_v\ge n\}$. The terminal conditions give $s\in S$ and $t\notin S$. Every potential in $S$ is at least $n$, and every potential outside $S$ is at most $n-1$. Thus every outgoing cut edge is full and every incoming edge is empty. Cut cancellation gives
\[
c(\delta^+(S))=\sum_{v\in S}\ex_f(v),
\qquad |c(\delta^+(S))-p|\le\frac{|S|}{8n}<1.
\]
Thus its integer capacity is $p$. Lemma~\ref{cor:mf-lower} gives the lower bound.
\end{proof}

\subsection{Completeness (why suitable potentials exist)}
Assume $\MaxFlow(s,t)=p\ge0$. Among flows of value $p$, minimize the quadratic energy normalized by the capacities,
\begin{equation}\label{eq:energy}
\Phi(f)=\sum_{(u,v):c_{uv}>0}\frac{f_{uv}^2}{2c_{uv}}.
\end{equation}
A minimizer $f^\star$ exists because the feasible set is nonempty, closed, and bounded. We call such a minimizer a minimum-energy flow.
The potentials below are the dual variables of this quadratic min-cost flow problem. Their relation to the flow is the classical optimality condition for separable convex costs~\cite{Roc84,AMO93}. We include its short residual-graph proof.

\paragraph{Residual optimality.}
For each edge $e=(u,v)$ with $c_e>0$, write $x_e=f_e^\star/c_e\in[0,1]$. If $f_e^\star<c_e$, include a forward residual edge $u\to v$ of length $x_e$; if $f_e^\star>0$, include a backward residual edge $v\to u$ of length $-x_e$. Each residual edge retains the identity of its original edge. In particular, antiparallel original edges may give parallel residual edges. The forward residual capacity is $c_e-f_e^\star$, and the backward residual capacity is $f_e^\star$.
\begin{lemma}[No negative residual cycle]\label{lem:no-neg}
The residual graph has no directed cycle of negative total length.
\end{lemma}
\begin{proof}
Suppose otherwise, and choose a simple directed residual cycle $C$ of length $L<0$. For every original edge $e$, let $\sigma_e\in\{-1,0,1\}$ be the number of times $C$ uses its forward residual edge minus the number of times it uses its backward residual edge. Increasing $f_e^\star$ by $\delta$ for a forward use and decreasing it by $\delta$ for a backward use gives the assignment
\[
f_e^\delta=f_e^\star+\delta\sigma_e.
\]
Since $C$ is a cycle, this change has zero net excess at every vertex, including $s$ and $t$. Thus $f^\delta$ has the same prescribed excess vector as $f^\star$.
Let $\rho>0$ be the minimum residual capacity of the edges of $C$. If $0<\delta<\rho$, each increase stays below its original capacity and each decrease stays nonnegative, so $f^\delta$ is capacity-feasible. The residual lengths give
\[
L=\sum_{e:c_e>0}\frac{\sigma_e f_e^\star}{c_e}.
\]
Put $A=\sum_{e:c_e>0}\sigma_e^2/c_e$. Since $L<0$, some $\sigma_e$ is nonzero, and hence $A>0$. Expanding the quadratic objective exactly yields
\[
\Phi(f^\delta)-\Phi(f^\star)
=\sum_{e:c_e>0}\frac{2\delta\sigma_e f_e^\star+\delta^2\sigma_e^2}{2c_e}
=\delta L+\frac{\delta^2}{2}A.
\]
Choose $0<\delta<\min\{\rho,-L/A\}$. Then the energy change is less than $\delta L/2<0$, contradicting the minimum-energy choice of $f^\star$.
\end{proof}

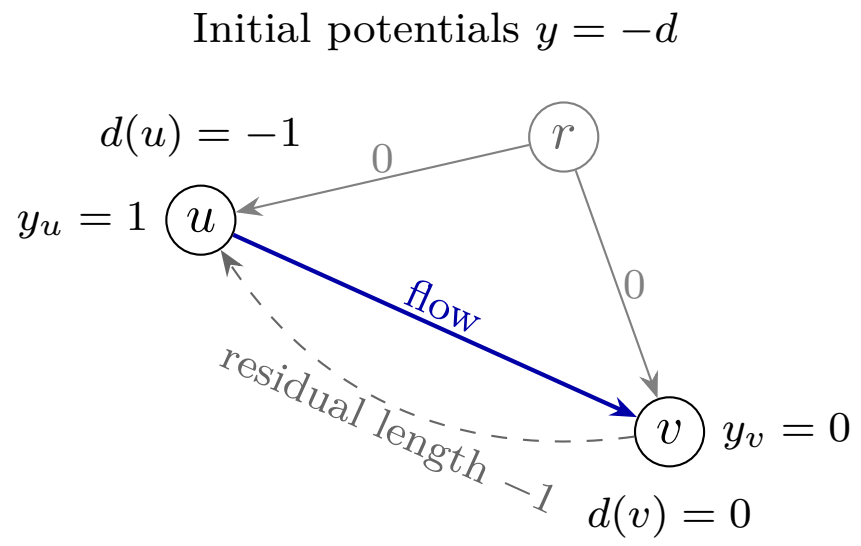
\begin{figure}[htbp]
\centering
\begin{tikzpicture}[x=1cm,y=1cm,
    vtx/.style={circle,draw,fill=white,inner sep=0pt,minimum size=13pt,font=\small},
    arc/.style={-{Stealth[length=5pt]}},
    lab/.style={font=\scriptsize,inner sep=1pt}]
  \node[lab] at (1.55,2.75) {Initial potentials $y=-d$};
  \node[vtx] (yu) at (0,1.5) {$u$};
  \node[vtx] (yv) at (3.1,.1) {$v$};
  \node[vtx,draw=gray,text=gray!80!black] (r) at (2.4,2.05) {$r$};
  \draw[arc,gray] (r) -- node[lab,above] {$0$} (yu);
  \draw[arc,gray] (r) -- node[lab,right] {$0$} (yv);
  \node[lab,above=4pt] at (yu.north) {$d(u)=-1$};
  \node[lab,below=4pt] at (yv.south) {$d(v)=0$};
  \node[lab,left=2pt] at (yu.west) {$y_u=1$};
  \node[lab,right=2pt] at (yv.east) {$y_v=0$};
  \draw[arc,thick,blue!65!black] (yu) -- node[lab,above,sloped] {flow} (yv);
  \draw[arc,dashed,gray!80!black] (yv) to[bend left=32]
    node[lab,below,sloped] {residual length $-1$} (yu);
\end{tikzpicture}
\caption{Shortest-path distances $d(u)=-1$ and $d(v)=0$ give potentials
$y_u=1$ and $y_v=0$, so positive flow goes from higher to lower potential.}
\label{fig:shortest-potentials}
\end{figure}

\paragraph{Recovering the flow from shortest-path potentials.}
Add a new root $r$ with zero-length edges to every vertex. This introduces no directed cycles, and every vertex is reachable from $r$. Lemma~\ref{lem:no-neg} therefore ensures that the shortest-path distance $d(v)$ is finite for every vertex. Set
\begin{equation}\label{eq:shortest-potentials}
y_v=-d(v).
\end{equation}
\noindent
A positive-flow edge $u\to v$ has a negative-length reverse edge. Following
it after a shortest path to $v$ gives $d(u)<d(v)$, so the initial potentials
$y=-d$ make flow go downhill (Figure~\ref{fig:shortest-potentials}).

\par\medskip\noindent\begin{minipage}{\linewidth}
\begin{lemma}[Clipped-potential representation]\label{lem:potentials}
Let $b$ be any feasible prescribed excess vector, and let $f^\star$
minimize the quadratic energy $\Phi$ among capacity-feasible edge
assignments with excess vector $b$. Then there exist potentials
$y\in[0,n-1]^V$ such that, for every original edge $(u,v)$,
\[
f_{uv}^\star=c_{uv}\clip(y_u-y_v,0,1).
\]
\end{lemma}
\end{minipage}\par
\begin{proof}
Shortest-path distances satisfy $d(w)\le d(v)+\ell$ for every residual edge $v\to w$ of length $\ell$. Now fix an original edge $(u,v)$ with $c_{uv}>0$ and put $x=f_{uv}^\star/c_{uv}\in[0,1]$. If $f_{uv}^\star<c_{uv}$, the forward residual edge $u\to v$ has length $x$. The distance inequality and $y_v=-d(v)$ give
\[
d(v)\le d(u)+x
\quad\Longrightarrow\quad
y_u-y_v=d(v)-d(u)\le x.
\]
If $f_{uv}^\star>0$, the backward residual edge $v\to u$ has length $-x$, giving
\[
d(u)\le d(v)-x
\quad\Longrightarrow\quad
y_u-y_v=d(v)-d(u)\ge x.
\]
There are three cases:
\begin{itemize}
\item If $f_{uv}^\star=0$, then $x=0$ and the forward edge exists. Thus $y_u-y_v\le0$, whose clipped value is $0=x$.
\item If $0<f_{uv}^\star<c_{uv}$, both residual edges exist. Their inequalities give $x\le y_u-y_v\le x$, so the drop and its clipped value both equal $x$.
\item If $f_{uv}^\star=c_{uv}$, then $x=1$ and the backward edge exists. Thus $y_u-y_v\ge1$, whose clipped value is $1=x$.
\end{itemize}
Multiplying the clipped value by $c_{uv}$ yields $f_{uv}^\star$ in every case. If $c_{uv}=0$, capacity feasibility gives $f_{uv}^\star=0$, and the claimed identity holds automatically.

For the range bound, the direct root edge $r\to v$ has length zero, so $d(v)\le0$. A shortest path can be chosen simple by deleting nonnegative cycles. After its first, zero-length root edge, it visits at most $n$ original vertices and hence uses at most $n-1$ residual edges. Every such edge has length at least $-1$. Consequently,
\[
-(n-1)\le d(v)\le0,
\qquad
0\le y_v=-d(v)\le n-1.
\]
The argument uses only energy minimality subject to fixed vertex excesses, so it applies to every fixed feasible excess vector $b$.
\end{proof}

\subsection{Rounding the potentials}\label{sec:rounding}
\begin{lemma}[Parameterized rounded potentials]\label{lem:certificate-parameters}
Consider an integer-capacity network on $r\le N$ vertices, where $N\ge2$
and $W'\ge1$ are integers and every capacity is at most $W'$. Put
\begin{equation}\label{eq:certificate-parameters}
Q(N,W')=16N^2W',\qquad
\mathcal Y_N=[0,N-1]\cup[N,2N-1],\qquad
\varepsilon_N=\frac1{8N}.
\end{equation}
Every feasible prescribed excess vector $b$ has a clipped-potential
representation whose potentials lie in $[0,r-1]$; rounding them down to
multiples of $1/Q(N,W')$ changes each excess by less than $\varepsilon_N$.
Each index uses $O(\log(NW'))$ bits.

For $s\ne t$ and an integer $p\ge0$, $\MaxFlow(s,t)=p$ if and only if
there are grid potentials $y_v\in\mathcal Y_N$ with $y_s\ge N$,
$y_t\le N-1$, and
$|\ex_y(v)-p(\one[v=s]-\one[v=t])|\le\varepsilon_N$ at every vertex.
\end{lemma}
\begin{proof}
Choose a minimum-energy flow $f$ with prescribed excess vector $b$. Lemma~\ref{lem:potentials} gives potentials $x_v\in[0,r-1]$ such that
$f_{uv}=c_{uv}\clip(x_u-x_v,0,1)$.
Put $Q=Q(N,W')$ and round each potential down to
$\widehat x_v=\lfloor Qx_v\rfloor/Q$. Write
$e_v=x_v-\widehat x_v\in[0,1/Q)$.
Since the rounding errors lie in the same interval,
\[
|(\widehat x_u-\widehat x_v)-(x_u-x_v)|
=|e_v-e_u|<\frac1Q.
\]
Clipping is $1$-Lipschitz: for all real $a,a'$, 
\[
|\clip(a,0,1)-\clip(a',0,1)|\le |a-a'|.
\]
This follows because clipping is nondecreasing, equals the identity on $[0,1]$, and is constant outside that interval.
Let $\widehat f_{uv}=c_{uv}\clip(\widehat x_u-\widehat x_v,0,1)$.
The clipped factor stays in $[0,1]$, so $\widehat f$ remains capacity-feasible, and
\[
|\widehat f_{uv}-f_{uv}|\le\frac{c_{uv}}Q.
\]
The inequality is strict when $c_{uv}>0$; zero-capacity edges have zero error.
At a vertex, we sum these errors over its incoming and outgoing edges. There are at most $r-1$ edges in each direction, each of capacity at most $W'$. Consequently,
\[
\begin{aligned}
|\ex_{\widehat f}(v)-b(v)|
&\le \sum_u\bigl(|\widehat f_{vu}-f_{vu}|+|\widehat f_{uv}-f_{uv}|\bigr)\\
&\le \frac1Q\sum_u(c_{vu}+c_{uv})
\le \frac{2(r-1)W'}Q\\
&\le\frac{2(N-1)W'}{16N^2W'}
=\frac{N-1}{8N^2}
<\frac1{8N}.
\end{aligned}
\]
Here $Q=16N^2W'$ and $r\le N$. The same estimate covers $r=1$, when there are no edges and the excess error is zero.

Now suppose $\MaxFlow(s,t)=p$, choose $f$ to be a minimum-energy maximum flow, and let $S$ be the source shore of a minimum cut. Cut cancellation gives
\[
0=c(\delta^+(S))-p
=\sum_{e\in\delta^+(S)}(c_e-f_e)
 +\sum_{e\in\delta^-(S)}f_e.
\]
Every summand is nonnegative, so each outgoing cut edge is saturated and each incoming cut edge is empty. Set $z_v=x_v+N\one[v\in S]$. Internal potential drops are unchanged. For an outgoing cut edge of positive capacity, its drop was already at least one and only increases; for an incoming cut edge of positive capacity, its drop was at most zero and only decreases. Thus the shift preserves every edge amount, including zero-capacity edges.
Vertices outside $S$ now have potentials in $[0,r-1]\subseteq[0,N-1]$, and vertices in $S$ have potentials in $[N,N+r-1]\subseteq[N,2N-1]$. Rounding these shifted potentials down to the same grid preserves their interval membership, because the interval endpoints are grid points. The error bound above applies to this representation as well, giving all required excess checks. Every resulting grid index is at most $(2N-1)Q$, so its encoding uses $O(\log(NW'))$ bits. The unshifted indices satisfy the same bound.

For the converse, apply the proof of Lemma~\ref{cor:encoded-value} with interval parameter $N$. Since the network has $r\le N$ vertices, the total excess error on any cut is at most $r/(8N)<1$, so the same capacity and integrality arguments give $\MaxFlow(s,t)=p$.
\end{proof}

We use the abbreviation $Q_0=Q(n,W)$ for the original network.
\begin{corollary}[Single-pair completeness]\label{cor:completeness}
If $\MaxFlow(s,t)=p\ge0$, the certificate has an accepting witness with
denominator $Q_0$ and $O(\log(nW))$ bits per potential.
\end{corollary}
\begin{proof}
Apply Lemma~\ref{lem:certificate-parameters} with $N=n$ and $W'=W$.
\end{proof}

\begin{proof}[Proof of Theorem~\ref{thm:intro-certificate}]
Fix $Q_0=16n^2W$ and encode each potential by the integer index $i_v=Q_0y_v$. Reject malformed or oversized fields, an incorrect number of indices, and claims outside $0\le p\le(n-1)W$. Check the interval and terminal conditions~\eqref{eq:cut-encoding} on these indices.
Initialize integer counters $E_v=0$. For each edge $(u,v)$, add the scaled amount
\[
F_{uv}=c_{uv}\clip(i_u-i_v,0,Q_0)=Q_0 f_{uv}
\]
to $E_u$ and subtract it from $E_v$. The final counters satisfy $E_v=Q_0\ex_f(v)$, so check
\[
|E_v-Q_0b(v)|\le Q_0/(8n).
\]
Indices are at most $(2n-1)Q_0$, and partial counters and targets have absolute value at most $(n-1)Q_0W$. Thus verification uses $O(n+m)$ arithmetic operations on $O(\log(nW))$-bit integers. Lemma~\ref{cor:encoded-value} proves soundness, and Corollary~\ref{cor:completeness} supplies an accepting certificate for every true claim.
\end{proof}

\section{Applications}\label{sec:applications}
\paragraph{Shared verification setup.}\label{par:shared-verification}
Streaming and Merlin--Arthur verification use the same polynomial and
capacity sketches. We describe their common checks here; the two models
differ in how the sketches are obtained and how many claims they serve.
Fix the denominator $Q_0=16n^2W$. We work over a prime field
$\mathbb F_P$, whose size is chosen below.
For a claimed integer $0\le p\le(n-1)W$, the proof supplies indices
$i_v=Q_0y_v$ satisfying the interval and terminal conditions
in~\eqref{eq:cut-encoding}. They define the capacity-feasible assignment
\begin{equation}\label{eq:streaming-assignment}
g_{uv}=\frac{c_{uv}}{Q_0}\clip(i_u-i_v,0,Q_0).
\end{equation}
The required excess checks are
\begin{equation}\label{eq:streaming-excess}
\left|Q_0\ex_g(v)-Q_0p(\one[v=s]-\one[v=t])\right|
\le\frac{Q_0}{8n}\qquad(v\in[n]).
\end{equation}
Lemma~\ref{cor:encoded-value} and Corollary~\ref{cor:completeness} show that
these checks certify the exact value and admit a proof for every true claim.

\paragraph{The polynomial and sketches.}
Let $\vartheta_u(X)$ be the Lagrange basis on $[n]$, and write
\[
\mathcal A_j^+(X)=\sum_u\vartheta_u(X)c_{uj},
\qquad
\mathcal A_j^-(X)=\sum_u\vartheta_u(X)c_{ju}.
\]
At a private point $\zeta$, put $\lambda_u=\vartheta_u(\zeta)$.
The two capacity sketches are
\begin{equation}\label{eq:ma-sketches}
\alpha_j=\mathcal A_j^+(\zeta)=\sum_u\lambda_uc_{uj},
\qquad
\beta_j=\mathcal A_j^-(\zeta)=\sum_u\lambda_uc_{ju}.
\end{equation}
They depend only on the graph and $\zeta$, and can therefore serve every
potential vector. For each vector define
\begin{equation}\label{eq:streaming-kernels}
\begin{aligned}
\mathcal K_j^+(X)&=\sum_u\vartheta_u(X)\clip(i_u-i_j,0,Q_0),\\
\mathcal K_j^-(X)&=\sum_u\vartheta_u(X)\clip(i_j-i_u,0,Q_0).
\end{aligned}
\end{equation}
Clipping is performed on integer indices before reduction into the field.
Define
\begin{equation}\label{eq:streaming-polynomial}
\mathcal E(X)=\sum_{j=1}^n
 \bigl(\mathcal A_j^+(X)\mathcal K_j^+(X)-
       \mathcal A_j^-(X)\mathcal K_j^-(X)\bigr).
\end{equation}
Each factor has degree at most $n-1$, so $\deg\mathcal E\le2n-2$.
At a vertex index $v$, the Lagrange basis selects vertex $v$, giving
$\mathcal E(v)=Q_0\ex_g(v)$ modulo $P$. At the private point, the
actual value is obtained from the stored sketches by
\[
\mathcal E(\zeta)=\sum_j
 \bigl(\alpha_j\mathcal K_j^+(\zeta)-\beta_j\mathcal K_j^-(\zeta)\bigr).
\]

\paragraph{Field convention.}
For a grid denominator $Q$, capacity bound $W$, and target error $n^{-c}$,
put $r_c=\lceil c\rceil$ and use a prime field of order $P$ with
\begin{equation}\label{eq:verification-field}
\begin{gathered}
M(Q,n,W,c)=\max\{4QnW,\,4n^{r_c+1}\},\\
M(Q,n,W,c)<P<2M(Q,n,W,c).
\end{gathered}
\end{equation}
Bertrand's postulate supplies $P$. This single choice separates all signed
scaled excesses and makes the $2n-1$ interpolation points required by
the degree bound distinct. The root bound gives
error below $n^{-c}$ for a degree-$(2n-2)$ check when each point has
probability at most $2/P$. For $Q=Q(n,W)=16n^2W$,
$\log P=O_c(\log(nW))$.

\paragraph{Checking the claimed excesses.}
Apply the \hyperref[par:polynomial-check]{polynomial evaluation check}
with $d=2n-2$. The prover supplies $2n-1$ claimed evaluations $h_i$.
The first $n$ are signed integers; the remaining $n-1$ are canonical field
elements. Let $H$ be their interpolating polynomial on $[2n-1]$ and
$\mu_i=\psi_i(\zeta)$ the corresponding weights. Put
$U_{\mathrm{ex}}=Q_0(n-1)W$ and check
\begin{equation}\label{eq:streaming-balance-check}
\begin{aligned}
|h_v|&\le U_{\mathrm{ex}},\\
\left|h_v-Q_0p(\one[v=s]-\one[v=t])\right|
 &\le Q_0/(8n)\qquad(v\in[n]),
\end{aligned}
\end{equation}
using signed integer arithmetic, together with
\begin{equation}\label{eq:streaming-polynomial-check}
H(\zeta)=\sum_{i=1}^{2n-1}\mu_i h_i=\mathcal E(\zeta).
\end{equation}
All range, interval, and encoding checks are deterministic. The entire
proof is fixed independently of $\zeta$.

If $H=\mathcal E$, then $h_v=Q_0\ex_g(v)$ modulo $P$ for each vertex.
Both integers lie in $[-U_{\mathrm{ex}},U_{\mathrm{ex}}]$, and $P>2U_{\mathrm{ex}}$,
so they are equal as signed integers. Thus~\eqref{eq:streaming-balance-check}
gives~\eqref{eq:streaming-excess}, and the accepted value is exact.
Consequently, any false claim passing the deterministic checks must have
$H\ne\mathcal E$ and is rejected with the error bound from the polynomial
check. For a true claim, the rounded certificate and actual evaluations
of $\mathcal E$ supply an accepting proof for every $\zeta$.

\paragraph{Constructing an accepting proof.}\label{sec:streaming-prover}
For a true claim, choose a minimum-energy maximum flow and obtain its potentials using Lemma~\ref{lem:potentials}. Lemma~\ref{lem:certificate-parameters} gives the rounded indices, and the actual evaluations of $\mathcal E$ at the prescribed interpolation points complete an accepting proof.

\subsection{Annotated streaming}\label{sec:streaming}
\paragraph{Model.}
In the online annotated streaming model~\cite{CCMT14,CGT20}, help may be interleaved with the input and depends only on the prefix seen so far. In our scheme, the verifier reads the input once using private randomness, then reads a proof from an untrusted prover and outputs a value or rejects. The proof may depend on the input but not on the private coins; the verifier sends no messages. The \emph{help cost} is the proof length in bits; the \emph{space cost} is the maximum working memory, including while reading the proof; their sum is the \emph{total cost}. Perfect completeness means every input has a proof accepted with its true value for every choice of coins. Soundness error at most $n^{-c}$ means every input and proof yields an incorrect value with probability at most $n^{-c}$.

\begin{samepage}
\begin{theorem}[Exact maximum-flow value in annotated streams]\label{thm:streaming}

Consider a turnstile stream of integer capacity updates for a directed network on $[n]$, with final capacities in $\{0,\ldots,W\}$ and specified terminals $s\ne t$, where $n\ge2$ and $W\ge1$. For every fixed $c>0$, $\MaxFlow(s,t)$ admits an online annotation scheme with perfect completeness and soundness error at most $n^{-c}$. The scheme uses $O_c(n\log(nW))$ bits of help and $O_c(n\log(nW))$ bits of working space, with all annotation following the input stream. For polynomially bounded $W$, its total cost is $\tO(n)$.
\end{theorem}
\end{samepage}
\begin{proof}
Use the \hyperref[par:shared-verification]{shared verification setup}.
The field and parameters $n,W,c$ are public. Before the input, privately
choose uniform $\zeta\in\mathbb F_P$, compute both sets of interpolation
weights, and initialize $\alpha_j=\beta_j=0$.
An update $(u,v,\Delta)$ adds $\Delta\lambda_u$ to $\alpha_v$ and
$\Delta\lambda_v$ to $\beta_u$. These are the linear sketches
in~\eqref{eq:ma-sketches}, so each update takes $O(1)$ field operations,
including deletions under the final-capacity promise.
After the input, read $p$, the $n$ potential indices, and the $2n-1$
evaluation-table entries. Reject malformed or oversized fields and perform
the shared interval, range, and excess checks. Compute the clipped
kernels from the stored indices and apply~\eqref{eq:streaming-polynomial-check}.
The shared correctness argument gives perfect completeness and error
at most $(2n-2)/P<n^{-c}$. No coins or intermediate check outcomes are
revealed, so the annotation remains independent of $\zeta$.

The weights, sketches, indices, and table require $O(n)$ words, and each
entry has $O_c(\log(nW))$ bits. Help and space are therefore each
$O_c(n\log(nW))$ bits. Direct evaluation of all kernels takes $O(n^2)$
field operations after the input; the model charges only help and space, not verification time.
Lemma~\ref{lem:ma-clipped} reduces this work to $O(n\log n)$ comparisons
and $O(n)$ field operations without increasing the space bound.
\end{proof}

\begin{corollary}[Optimality for online annotation]\label{cor:streaming-optimality}
Every bounded-error online annotation scheme for exact maximum-flow value with help cost $h_{\mathrm{ann}}\ge1$ and space cost $v_{\mathrm{mem}}$ satisfies $h_{\mathrm{ann}}v_{\mathrm{mem}}=\Omega(n^2)$, and hence $h_{\mathrm{ann}}+v_{\mathrm{mem}}=\Omega(n)$. The total cost of Theorem~\ref{thm:streaming} is optimal up to $O_c(\log(nW))$, and its help--space product is optimal up to $O_c(\log^2(nW))$.
\end{corollary}
\begin{proof}
For a bipartite graph with public equal parts $V_L,V_R$, direct each streamed edge from $V_L$ to $V_R$ at unit capacity. Insert fixed unit edges $s\to V_L$ and $V_R\to t$ before the input. Integrality makes the maximum-flow value equal the maximum matching size, so a perfect matching exists exactly when that value is $|V_L|$. This reduction uses $n+2$ vertices and preserves online annotation and its costs. The product bound for online bipartite perfect matching~\cite{CCMT14} transfers; the sum bound follows by the arithmetic--geometric mean inequality.
\end{proof}

\begin{samepage}
\begin{corollary}[Certificate-first streaming]\label{cor:cert-first}

Let $n\ge2$, $W\ge1$, and terminals $s\ne t$ be public. Consider an arbitrary-order stream of directed edges with nonnegative integer capacities, with total capacity at most $W$ on each ordered pair. The predicate $\MaxFlow(s,t)=p$ admits a deterministic one-pass verifier in the certificate-first model of~\cite{DFPR25}, with perfect completeness and soundness. The certificate is independent of the stream order and is available in read-only memory before the stream. Certificate size and working space are each $O(n\log(nW))$ bits.
\end{corollary}
\end{samepage}
\begin{proof}
Run the verifier from the proof of Theorem~\ref{thm:intro-certificate}, reading the $n$ potential indices from the certificate and accumulating the excess counters in one pass over the stream. The nonnegative updates keep every partial counter within $(n-1)Q_0W$ in absolute value, where $Q_0=16n^2W$, so the certificate and counters each use $O(n\log(nW))$ bits. Correctness follows from the single-pair certificate and is independent of the stream order.
\end{proof}

\paragraph{Optimality for certificate-first streaming.}
The certificate size plus working space must be $\Omega(n)$, already for unit capacities. Apply the nondeterministic disjointness reduction in the proof of~\cite{DFPR25} to its bipartite matching instance, with two parts of size $N$. Orient the edges from left to right and add unit edges from a new source to the left part and from the right part to a new sink. The maximum-flow value is $N$ exactly when the instance has a perfect matching, so their $\Omega(N)$ total-cost lower bound transfers to exact-value certification on $2N+2$ vertices. Corollary~\ref{cor:cert-first} is therefore optimal in total cost up to logarithmic factors.

\begin{remark}[Help--space tradeoff]\label{rem:streaming-tradeoff}
For $1\le\ell\le n$, the online scheme admits $\tO(n^2/\ell)$ help
and $\tO(\ell)$ space, optimal in help--space product up to logarithmic
factors. Put $k=\lceil n/\ell\rceil$ and group the columns into $k$
blocks of size at most $\ell$. Bivariate reshaping~\cite{CG19,CGT20}
yields a polynomial of bidegree $(2n-2,2k-2)$ evaluable from $2\ell$
capacity sketches. The prover supplies a $(2n-1)\times(2k-1)$ table;
the upper-left $n\times k$ subtable has signed entries whose row sums
are the scaled vertex excesses. The verifier processes potentials
blockwise in $O(\ell)$ words, using private fingerprints to check
repeated copies. With independent private points and a prime
$8M<P<16M$, where $M$ is from~\eqref{eq:verification-field},
the excess and polynomial checks have perfect completeness and
soundness error below $n^{-c}$.
\end{remark}

\subsection{Broadcast proof-labeling schemes}\label{sec:local-max}
The certificate of Section~\ref{sec:single-certificate} also gives a one-round broadcast proof-labeling scheme. Each vertex checks its own excess, with both endpoints deriving an edge amount from their two potentials.

\paragraph{Model.}
We use one-round proof-labeling schemes~\cite{KKP10}. The underlying communication graph is connected, and vertices have distinct nonnegative integer identifiers of $O(\log n)$ bits. For a directed network, communication runs both ways along each edge. Each vertex knows the directions and capacities of its incident edges and its terminal role. The data at the two endpoints of an edge agree. The claimed value $p$ is part of the local input at every vertex. A prover supplies one label per vertex; each vertex broadcasts that same label to all its neighbors, and decides from its input and the labels at itself and its neighbors. Every true instance has labels making all vertices accept, and every labeling of a false instance makes some vertex reject. We measure the maximum label length on true instances; this also bounds broadcast message length.

\begin{corollary}[Local certification of maximum-flow values]\label{cor:local-max}

For directed networks with nonnegative integer capacities, distinct terminals $s,t$, and every integer $p\ge0$, the predicate $\MaxFlow(s,t)=p$ admits a broadcast proof-labeling scheme with $O(\log n+\log(p+1))$ bits per vertex. The same bound holds for undirected networks, by treating each edge as two opposite edges of the same capacity. For $p\ge n^{\Omega(1)}$, the bound is optimal up to constant factors.
\end{corollary}
\begin{proof}
Set $\widehat c_e=\min\{c_e,p+1\}$. The capped network has value $p$ exactly when the original does. Indeed, a cut changed by capping has capped capacity at least $p+1$, whereas every cut of capacity at most $p$ is unchanged. This also covers $p=0$.
Each label contains a common integer $N_{\rm id}\ge2$ and one grid
index $i_v$. Neighbors check agreement on $N_{\rm id}$, and every vertex
checks $\operatorname{id}(v)<N_{\rm id}$; connectedness and distinct
identifiers imply $N_{\rm id}\ge n$.
Apply Lemma~\ref{lem:certificate-parameters} to the capped network with
$(N,W')=(N_{\rm id},p+1)$. Each vertex checks its interval, terminal
condition, and excess tolerance from that lemma using its own and its
neighbors' indices. The lemma proves soundness and completeness.
For an honest certificate take
$N_{\rm id}=\max\{2,1+\max_v\operatorname{id}(v)\}$, which has
$O(\log n)$ bits; each grid index then has
$O(\log n+\log(p+1))$ bits.

\emph{Optimality.} We extend the crossing argument of~\cite{PP22} to every $p\ge1$. Fix $n\ge6$ and vertex identifiers. For $a,b\in\{0,\ldots,p\}$, use the six-vertex network in Figure~\ref{fig:broadcast-crossing}, padded to $n$ vertices by a path of zero-capacity edges at $t$.
\begin{figure}[H]
\centering
\vspace{-4pt}
\begin{tikzpicture}[>=Stealth, every node/.style={font=\small}, vertex/.style={circle,draw,inner sep=1.5pt,minimum size=0.51cm}, edgelabel/.style={fill=white,inner sep=1pt,font=\small}]
  \node[vertex] (s) at (0,0) {$s$};
  \node[vertex] (x1) at (1.35,0.65) {$x_1$};
  \node[vertex] (x2) at (1.35,-0.65) {$x_2$};
  \node[vertex] (y1) at (3.10,0.65) {$y_1$};
  \node[vertex] (y2) at (3.10,-0.65) {$y_2$};
  \node[vertex] (t) at (4.45,0) {$t$};
  \draw[->] (s) -- node[edgelabel,above left,pos=.54] {$a$} (x1);
  \draw[->] (s) -- node[edgelabel,below left,pos=.54] {$p-a$} (x2);
  \draw[->] (x1) -- node[edgelabel,above,pos=.32] {$p$} (y1);
  \draw[->] (x2) -- node[edgelabel,below,pos=.32] {$p$} (y2);
  \draw[->] (y1) -- node[edgelabel,above right,pos=.46] {$b$} (t);
  \draw[->] (y2) -- node[edgelabel,below right,pos=.46] {$p-b$} (t);
  \draw[densely dashed,gray] (2.23,-0.95) -- (2.23,0.95);
  \node[font=\footnotesize] at (0.75,-1.10) {left: $a$};
  \node[font=\footnotesize] at (3.80,-1.10) {right: $b$};
\end{tikzpicture}
\vspace{-8pt}
\caption{Six-vertex network for the broadcast proof-labeling lower bound.}
\label{fig:broadcast-crossing}
\vspace{-8pt}
\end{figure}
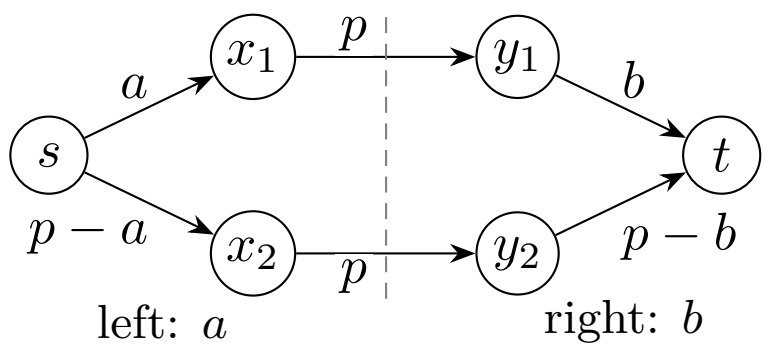
The two $s$--$t$ paths carry $\min\{a,b\}$ and $\min\{p-a,p-b\}$, so the value is $p-|a-b|$. Fix an accepting labeling for each instance with $a=b$. If two values $a\ne a'$ gave the same labels at $x_1,x_2,y_1,y_2$, take $s,x_1,x_2$ with their capacities and labels from instance $a$, and the other vertices from instance $a'$. Every vertex then sees the input and labels of an accepted instance, so all accept, although the value is $p-|a-a'|\ne p$. Hence the four-label tuple takes at least $p+1$ distinct values, so the largest of these four labels requires $\Omega(\log(p+1))$ bits. The graph has maximum degree three, and the same argument applies with undirected edges. For $p\ge n^{\Omega(1)}$, this matches the upper bound.
\end{proof}

\noindent\begin{minipage}{\linewidth}
\paragraph{Comparison with other certification tasks.}
Here only the claimed value $p$ is given, not a flow. Thus, the one-bit
scheme of Emek and Gil~\cite{EG20} for verifying a supplied maximum flow
does not apply. Their $O(\log n+\log W)$-bit scheme certifies only an
upper bound, whereas we certify $\MaxFlow(s,t)=p$. This is the $p$-MF
task of~\cite{PP22}, distinct from their supplied-flow formulation.
\end{minipage}

\subsection{Merlin--Arthur verification}\label{sec:ma}
\paragraph{Model.}
In the one-message Merlin--Arthur model, Merlin fixes one proof
independently of Arthur's private coins. Arthur then accepts or
rejects the entire collection of claims. Perfect completeness requires
some proof to be accepted with probability one when every claim is true;
soundness error at most $n^{-c}$ requires every proof to be accepted with
probability at most $n^{-c}$ when any claim is false. Verification time
includes reading the proof.

We evaluate the shared polynomial check quickly and reuse its graph
sketches across all pairs.

\begin{samepage}
\begin{theorem}[Amortized Merlin--Arthur verification]\label{thm:ma}

Let $G$ be a directed graph on $[n]$, $n\ge2$, with $m$ nonzero-capacity edges. For any public list of $1\le q\le n(n-1)$ ordered terminal pairs with claimed integer maximum-flow values, and every fixed $c>0$, there is a one-message Merlin--Arthur protocol with perfect completeness and soundness error at most $n^{-c}$. The protocol uses $O_c(qn\log(nW))$ proof bits, $\tO(m+qn)$ verification time, and $\tO(n)$ bits of working space. For a capacity-matrix input, verification takes $\tO(n^2+qn)$ time. In particular, directed All-Pairs Max-Flow admits $\tO(n^3)$ verification time with $O_c(n^3\log n)$ proof bits, independently of $\omega$.
\end{theorem}
\end{samepage}

Use the \hyperref[par:shared-verification]{shared verification setup}.
Its capacity sketches depend only on the graph and the private point.
The additional task is to evaluate the clipped kernels quickly for each
potential vector, which the next lemma provides.

\begin{lemma}[Sorted clipped-sum evaluation]\label{lem:ma-clipped}
Given the integer indices $i_1,\ldots,i_n$, the weights
$\lambda_u=\vartheta_u(\zeta)$, and the two capacity sketches
$\alpha_j=\mathcal A_j^+(\zeta)$ and
$\beta_j=\mathcal A_j^-(\zeta)$, the true value $\mathcal E(\zeta)$ can be
computed using $O(n\log n)$ integer comparisons, $O(n)$ field operations,
and $O(n)$ additional words.
\end{lemma}
\begin{proof}
Sort the vertices as $\sigma(1),\ldots,\sigma(n)$ by nondecreasing index.
Form the weighted prefix sums
\[
S_k=\sum_{r\le k}\lambda_{\sigma(r)},\qquad
T_k=\sum_{r\le k}\lambda_{\sigma(r)}i_{\sigma(r)},\qquad S_0=T_0=0.
\]
For each vertex $j$, define intervals in this order by
\[
\begin{aligned}
\mathcal U_j&=\{u:i_j<i_u<i_j+Q_0\},&
\mathcal V_j&=\{u:i_u\ge i_j+Q_0\},\\
\mathcal L_j&=\{u:i_j-Q_0<i_u<i_j\},&
\mathcal B_j&=\{u:i_u\le i_j-Q_0\}.
\end{aligned}
\]
The clipping formulas give
\begin{equation}\label{eq:ma-clipped-sums}
\begin{aligned}
\mathcal K_j^+(\zeta)
 &=\sum_{u\in\mathcal U_j}\lambda_ui_u
   -i_j\sum_{u\in\mathcal U_j}\lambda_u
   +Q_0\sum_{u\in\mathcal V_j}\lambda_u,\\
\mathcal K_j^-(\zeta)
 &=i_j\sum_{u\in\mathcal L_j}\lambda_u
   -\sum_{u\in\mathcal L_j}\lambda_ui_u
   +Q_0\sum_{u\in\mathcal B_j}\lambda_u.
\end{aligned}
\end{equation}
The strict and weak inequalities handle ties and both breakpoints exactly.
After sorting, all interval endpoints can be found either by binary search
in $O(n\log n)$ comparisons or by moving pointers in $O(n)$ comparisons.
The prefix arrays then evaluate each sum in constant time, giving all $2n$
kernel values in $O(n)$ field operations; their dot products with $\alpha$
and $\beta$ give $\mathcal E(\zeta)$ in another $O(n)$ operations.
\end{proof}

\begin{proof}[Proof of Theorem~\ref{thm:ma}]
With $M=M(Q_0,n,W,c)$ from~\eqref{eq:verification-field}, Merlin supplies a prime $M<P<2M$, whose range and primality Arthur checks deterministically~\cite{AKS04}. Since $\log P=O_c(\log(nW))$, the header and its verification add only logarithmic bits and polylogarithmic time; subsequent arithmetic is over $\mathbb F_P$.
Arthur draws $\lceil\log_2P\rceil$ private random bits and reduces the
resulting integer modulo $P$ to obtain $\zeta\in\mathbb F_P$; each point
has probability at most $2/P$.
He computes the shared interpolation weights $\lambda_u,\mu_i$ in
$\tO(n)$ time and the capacity sketches~\eqref{eq:ma-sketches} in one
graph pass using $O(m+n)$ field operations.
These weights and sketches serve every query.

\paragraph{Query blocks and checks.}
For each query $a=(s,t)$, Arthur reads the potential indices and
evaluation table in the shared format, rejecting malformed fields or
fields exceeding their prescribed encoding limits immediately.
He checks $0\le p_a\le(n-1)W$, the interval and terminal
conditions~\eqref{eq:cut-encoding}, and the integer range and excess
conditions~\eqref{eq:streaming-balance-check}.
Writing $H_a$ for the table's interpolant, he computes
$\mathcal E_a(\zeta)$ by Lemma~\ref{lem:ma-clipped} and performs
the polynomial check~\eqref{eq:streaming-polynomial-check}.
Arthur discards each completed query's data and accepts exactly when
every block passes and the proof has the prescribed length.
The entire message, including the prime and all query blocks, is fixed
independently of $\zeta$.

\paragraph{Perfect completeness.}
For each true claim, use the accepting proof from the shared setup.
It gives $H_a=\mathcal E_a$ and every integer excess check, so Arthur
accepts for every $\zeta$.

\paragraph{Soundness.}
Fix the graph, the claims, and the entire proof, including its verified
prime. If any claim is false and all deterministic checks pass, the
shared correctness argument gives $H_a\ne\mathcal E_a$ for at least
one pair. Apply the polynomial evaluation check to this fixed pair,
with degree bound $2n-2$ and point probabilities at most $2/P$:
\[
\Pr[\mathrm{accept}]\le\frac{2(2n-2)}{P}<n^{-c}.
\]
The same point serves all queries without a factor of $q$, as explained
in the shared polynomial check.

\paragraph{Costs.}
For each query, sorting uses $O(n\log n)$ comparisons, and the clipped
sums and table check use $O(n)$ field operations. The first $n$ table
entries take $O(n)$ integer checks. Each field element or index occupies
$O_c(\log(nW))$ bits. Thus the total verification time is
$\tO(m+qn)$ and the proof length is $O_c(qn\log(nW))$ bits.
The weights, graph sketches, and one query block use $\tO(n)$ bits of
working space. Reading a dense capacity matrix costs $O(n^2)$, and
$q=n(n-1)$ gives the all-pairs bounds.
\end{proof}

\subsection{Deterministic all-pairs verification}\label{sec:batch}
\paragraph{Model.}
The input is a directed graph's integer capacity matrix, with
polynomially bounded entries. The algorithm guesses a value and
rounded potentials for each ordered terminal pair, then verifies
them deterministically. Failed checks return ``don't know'';
running time includes reading the guesses.

We batch the max-flow vertex checks in~\eqref{eq:approx-ex} against a fixed capacity matrix.
Let $\mathcal Q$ be a set of $q$ ordered pairs of distinct vertices, each with its own potential vector and pseudoflow; for All-Pairs Max-Flow, $\mathcal Q$ consists of all $n(n-1)$ ordered pairs.
Index pairs by $a=(s,t)\in \mathcal Q$, and write $b_a=p_a(\evec_s-\evec_t)$ for the claimed vertex excesses. For every pair $a$ and vertex $v$, compute
\begin{align}
\operatorname{out}_a(v)&=\sum_u c_{vu}\clip(y_v^a-y_u^a,0,1),\label{eq:out}\\
\operatorname{in}_a(v)&=\sum_u c_{uv}\clip(y_u^a-y_v^a,0,1).\label{eq:in}
\end{align}

\subsubsection{Clipped sums in sorted order}
Fix pair $a$ and sort vertices by $y_u^a$. For fixed $v$,
\[
\clip(y_v^a-y_u^a,0,1)=
\begin{cases}
1,& y_u^a\le y_v^a-1,\\
y_v^a-y_u^a,&y_v^a-1<y_u^a<y_v^a,\\
0,& y_u^a\ge y_v^a.
\end{cases}
\]
The incoming factor has thresholds $y_v^a$ and $y_v^a+1$. A full interval needs only two aggregates: total capacity and potential-weighted capacity (Figure~\ref{fig:blocking}).

\subsubsection{Buckets and four matrix products}
Fix an integer bucket size $1\le B\le n$ and partition each sorted list into consecutive buckets of size at most $B$. Across all pair-bucket combinations, the number of rows is
\[
r_{\rm rows}=q\lceil n/B\rceil.
\]
Write $i_u^a=Q_0y_u^a$ for the integer grid indices and define $r_{\rm rows}\times n$ integer matrices
\[
\mathbf X_{(a,I),u}=\one[u\in I],\qquad \mathbf Z_{(a,I),u}=i_u^a\one[u\in I].
\]
Let $\mathbf A$ be the capacity matrix. Compute
\begin{equation}\label{eq:four-products}
\mathbf X\mathbf A,\qquad \mathbf X\mathbf A^\top,\qquad \mathbf Z\mathbf A,\qquad \mathbf Z\mathbf A^\top.
\end{equation}
For pair-bucket $(a,I)$ and vertex $v$, the entries of these four products are, respectively,
\[
\sum_{u\in I}c_{uv},\qquad\sum_{u\in I}c_{vu},\qquad
\sum_{u\in I}i_u^ac_{uv},\qquad\sum_{u\in I}i_u^ac_{vu},
\]
namely the incoming and outgoing capacities and their counterparts weighted by the scaled potentials. Dividing the last two sums by $Q_0$ gives the potential-weighted capacity sums used above. Prefix sums over the buckets of each pair evaluate unions of whole buckets in constant time, and binary search locates the breakpoints. In each direction the two breakpoints lie in at most two buckets, which are scanned directly, at cost $O(B)$ per pair-vertex (Figure~\ref{fig:blocking}).
\begin{lemma}[Batch verification time]\label{lem:batch}
Write $\MM_d(r_{\rm rows})$ for the time to multiply an $r_{\rm rows}\times d$ matrix by a $d\times d$ matrix. For every integer bucket size $1\le B\le n$, all excess tests for the pairs in $\mathcal Q$ can be performed in
\[
\tO\bigl(\MM_n(q\lceil n/B\rceil)+qnB\bigr)
\]
time.
\end{lemma}
\begin{proof}
Sorting the $q$ potential vectors costs $\tO(qn)$. The four products cost $O(\MM_n(r_{\rm rows}))$. Their outputs have $r_{\rm rows}n$ entries, so the prefix sums cost $O(r_{\rm rows}n)$, which is $O(\MM_n(r_{\rm rows}))$ because a product must write its output. There are $qn$ pair-vertex combinations, each scanning at most four buckets of size at most $B$, for $O(qnB)$ additional time.
\end{proof}
\begin{corollary}[All-pairs excess-test time]\label{cor:all-batch}

For all ordered terminal pairs, the excess tests take $n^{\gamma+o(1)}$ time, where
\[
\gamma=\min_{0\le\beta\le1}\max\{\omega(3-\beta,1,1),\,3+\beta\}.
\]
\end{corollary}
\begin{proof}
Set $B=\lceil n^\beta\rceil$. Here $r_{\rm rows}\le n^2\lceil n/B\rceil=O(n^{3-\beta})$, so multiplying the full rectangular matrices costs $n^{\omega(3-\beta,1,1)+o(1)}$, while scans cost $n^{3+\beta+o(1)}$. Optimize over $0\le\beta\le1$. The standard bound $\omega(3-\beta,1,1)\le1-\beta+\omega(2,1,1)$ also gives $\gamma\le(4+\omega(2,1,1))/2$. If $\omega=2$, then $\omega(3-\beta,1,1)=4-\beta$, so $\gamma=7/2$.
\end{proof}

\begin{algorithm}[t]
\caption{Nondeterministic algorithm for directed All-Pairs Max-Flow}\label{alg:all-pairs}
\noindent\textbf{Input:} The capacity matrix $\mathbf A$ of a directed graph on $n$ vertices.
\begin{enumerate}[label=\arabic*:,leftmargin=2.3em,itemsep=1pt,parsep=0pt,topsep=4pt]
\item Guess, for every ordered pair $a=(s,t)$, an integer $0\le p_a\le\min\{d_c^+(s),d_c^-(t)\}$ and $n$ grid indices with denominator $Q_0$. Check that the resulting potentials satisfy~\eqref{eq:cut-encoding}. Return ``don't know'' on malformed fields or failed checks.
\item Sort each $y^a$ and cut its sorted order into buckets of size $B=\lceil n^\beta\rceil$, choosing $\beta$ as in Corollary~\ref{cor:all-batch}.
\item Compute $\mathbf X\mathbf A$, $\mathbf X\mathbf A^\top$, $\mathbf Z\mathbf A$, $\mathbf Z\mathbf A^\top$ by rectangular matrix multiplication, and their prefix sums over buckets.
\item For every pair $a$ and vertex $v$, compute $\operatorname{out}_a(v)$ and $\operatorname{in}_a(v)$ from whole-bucket aggregates and scans of at most four breakpoint buckets. Return ``don't know'' if $|\operatorname{out}_a(v)-\operatorname{in}_a(v)-b_a(v)|>1/(8n)$.
\item \textbf{return} the matrix $(p_a)$
\end{enumerate}
\end{algorithm}

\begin{corollary}[Global all-pairs verification]\label{cor:global-all-pairs}

For directed graphs, Algorithm~\ref{alg:all-pairs} is a nondeterministic algorithm for the ordered-pair maximum-flow matrix. Its running time is
\[
n^{\gamma+o(1)},\qquad \gamma\text{ as in Corollary~\ref{cor:all-batch}}.
\]
\end{corollary}
\begin{proof}\label{par:global-correctness}
Lemma~\ref{cor:encoded-value} and Corollary~\ref{cor:completeness} show that accepted pair values are exact and that the true matrix admits an accepting guess. The fixed format uses $O(n^3\log n)$ bits, with malformed fields returning ``don't know'', so the bound in Corollary~\ref{cor:all-batch} covers both reading the guess and checking every vertex. Using the bounds $\omega(2,1,1)<3.250385$ and $\omega(2.5,1,1)<3.720468$ from~\cite[Table~1]{WXXZ24}, interpolate with $\beta=0.6137$. Since $3-\beta=2.3863$, we have
\[
\omega(3-\beta,1,1)
 < (1-0.7726)\cdot 3.250385+0.7726\cdot 3.720468
 <3.6137=3+\beta.
\]
Consequently, $\gamma\le3.6137<3.614$. For polynomially bounded integer capacities, the matrix entries and products have $O(\log n)$ bits, so exact integer matrix multiplication incurs only polylogarithmic overhead.
\end{proof}

\section{Open problems}\label{sec:open}
\begin{enumerate}
\item \textbf{Faster Merlin--Arthur verification.} Does directed All-Pairs Connectivity admit a near-quadratic Merlin--Arthur verifier, as 4-Clique does~\cite{ACJRW22}?
\item \textbf{Faster deterministic verification.} Can All-Pairs Max-Flow be verified deterministically in $n^{\omega(2,1,1)+o(1)}$ time, matching the time for verifying upper-bound cuts and giving $n^{3+o(1)}$ time if $\omega=2$? Akmal~\cite{Akm24} asks for this improvement in the unit-capacity case, while Theorem~\ref{thm:ma} gives $\tO(n^3)$ randomized verification independently of $\omega$. For general capacities, our deterministic verifier takes $n^{3.5+o(1)}$ time when $\omega=2$ (Theorem~\ref{thm:intro-main}).
\item \textbf{Faster algorithms.} Does All-Pairs Max-Flow admit an $O(n^{4-\varepsilon})$-time algorithm on dense directed graphs?
\end{enumerate}

\noindent\textbf{AI disclosure.} The writeup and ideas throughout this paper were developed jointly by the author, ChatGPT (OpenAI), and Claude (Anthropic). The author is fully responsible for the integrity, accuracy, and originality of the paper.

\begingroup
\normalsize
\newcommand{\doi}[1]{\begingroup\def\UrlBreaks{\do\/\do\.\do\-}\def\UrlBigBreaks{\do\/}\href{https://doi.org/#1}{\textcolor{blue}{\nolinkurl{doi:#1}}}\endgroup}

\endgroup

\end{document}